\documentclass[letterpaper,12pt]{article}
\usepackage[letterpaper, margin=1in]{geometry}
\usepackage[utf8]{inputenc}
\usepackage{verbatim,array,multicol,palatino,amsfonts,amsmath, subfig, psfrag}
\usepackage{chicago}
\usepackage{color}
\usepackage[pdftex]{graphicx}
\usepackage{epstopdf}
\usepackage{stackengine}
\usepackage{latexsym,amssymb,epsfig,graphics, mathtools,amsthm, dcolumn,titlesec, authblk}
\newtheorem{proposition}{Proposition}

\usepackage{setspace}
\usepackage[normalem]{ulem}

\DeclareMathOperator*{\argmin}{arg\,min}
\titleformat*{\section}{\large\bfseries}

\title{ \bf {Noncrossing simultaneous Bayesian quantile curve fitting}}
\date{}
\author{T. Rodrigues\footnote{School of Mathematics and Statistics, University of New South Wales, Sydney 2052 Australia.}\:\,\footnote{CAPES Foundation, Ministry of Education of Brazil, Bras\'{i}lia - DF 70040-020, Brazil}\:\,
 J.-L. Dortet-Bernadet\footnote{Institut de Recherche Math\'ematique Avanc\'ee, UMR 7501 CNRS, Universit\'e de Strasbourg, Strasbourg, France.}\:\:  and Y. Fan$^{*}$ {\footnote{Communicating Author: {\tt Y.Fan@unsw.edu.au}}}}

\begin{document}

\maketitle

%\vspace{-0.5cm}

\begin{abstract}
Bayesian simultaneous estimation of nonparametric quantile curves is a challenging problem, requiring a flexible and robust data model whilst satisfying the monotonicity or noncrossing constraints on the quantiles. 
{This paper presents the use of the pyramid quantile regression method
 %recently proposed by the authors for simultaneous quantile regression,
 in the spline regression setting.} 
In high dimensional problems, the choice of the pyramid locations becomes crucial for a robust parameter estimation. 
In this work we derive the optimal  {pyramid locations which then allows us to propose an efficient} adaptive block-update MCMC scheme for posterior computation.
 %\textcolor{magenta}{able to estimate quantile curves which automatically satisfy the ordering constraints.} 
Simulation studies show the proposed method provides estimates with significantly smaller errors and better empirical coverage probability when compared to existing alternative approaches. 
We illustrate the method with three real applications.
\\

\noindent {\bf Keywords:} Bayesian quantile pyramid; Simultaneous quantile regression; B-Splines;  O'Sullivan penalised splines; Nonparametric quantile regression.                                                                       
\end{abstract}

\section{Introduction} \label{sec:intro}
Quantile regression models (QR; \citeNP{KoenkerBasset1978}) provide a comprehensive description of the conditional distribution of the response variable by capturing the effect of covariates at different quantile levels. It is a robust alternative to ordinary mean regression and has been embraced by a variety of fields, including {biology}, economics, environmental sciences, medicine and ecology (\shortciteNP{bookQR}, \shortciteNP{ReichFuentesDunson2011}, \citeNP{Portnoy03}, \shortciteNP{ecol99}). For modelling the $\tau$-th conditional quantile curve of a response variable $Y$ given the value $X=x$ of some covariate $X$ as $Q_\tau(Y|x)=f_\tau(x)$, $\tau \in (0,1)$, consider
\begin{eqnarray*}
Y|{X= x} & \sim f_\tau(x) + \epsilon,
\end{eqnarray*}
where the error variable $\epsilon$ satisfies $Q_{\tau} (\epsilon|{ x}) \equiv \inf\{a: P(\epsilon \leq a |X={x}) \geq \tau\} = 0$, and the function $f_\tau(x)$ changes with $\tau$ and describes  the relationship between $X$ and $Y$.
In this article, we are interested in the case where the curve $f_{\tau}$ is 
modeled with spline functions of a given degree, $P\geq 1$, so that,
\begin{equation}\label{eqn:spline}
f_{\tau}(x) =  \alpha_0 + \sum_{j=1}^P \alpha_jx^j + \sum_{k=1}^K \eta_k(x-\gamma_k)_+^P
\end{equation}
where $z_+=\max(0,z)$ and where $\gamma_k, k=1,\ldots,K $, represent the locations of  $K$ knot points (see \citeNP{hastie+t90}). 
Typically,  the degree $P$ is set to 3, since  cubic splines are known to approximate locally smooth functions sufficiently well.

Frequentist literature on quantile regression is largely based on the seminal work by \citeN{KoenkerBasset1978}, in which the noise distribution is left unspecified and quantiles are estimated by solving the linear optimisation problem
\begin{eqnarray}
\widehat{Q}_\tau(Y|x) = \argmin_{ f_\tau(x) } \sum_{i=1}^n \rho_{\tau}(y_i-f_\tau(x_i)) \, ,
\label{eq:qrkoenker}
\end{eqnarray}
where $\rho_{\tau}(\cdot)$ is an asymmetric loss function given by $\rho_{\tau}(\epsilon)=\tau \epsilon $ if $\epsilon\geq 0$, and $\rho_{\tau}(\epsilon)=(\tau-1) \epsilon$ otherwise. In the spline fitting context, a penalty term is added to \eqref{eq:qrkoenker} in order to restrict the class of functions $f_\tau(x)$ to be sufficiently smooth. However, the simplicity of linear programming is jeopardised when the classical quadratic penalty, $L_2=\int (f''(x))^2dx$, is added to this objective function (\shortciteNP{BOSCH95}). 
\shortciteN{koenker94} advocate the use of a linear norm under the total variation roughness penalty, $L_1=\int|(f''(x))|dx$, nevertheless one becomes harshly delimited by the space of piecewise linear fits. On the other hand, under the scope of linear programming, extension to quadratic splines is still possible when 
using the $L_\infty$-norm, $L_\infty = \text{sup}|f''(x)|$, see {\it e.g.} \shortciteN{koenker94} or \citeN{He1999}.  
In the Bayesian framework, the need for a likelihood specification was initially accommodated by the asymmetric Laplace error model introduced by \citeN{YuMoyeed2001}, based on its analogy to the minimisation problem \eqref{eq:qrkoenker}. Smoothing cubic splines can be easily incorporated considering a prior density for $f_\tau(x)$ proportional to $\text{exp}\{-\frac{1}{2}\lambda \int f''(x)^2dx\}$ (see \shortciteNP{Thompson2010}). 

However, one common issue facing these approaches is that quantiles at different levels have to be fitted singly for each $\tau$ and final estimates may cross, ie. estimates may not respect the monotonicity of the quantile function. In the context of nonparametric regression, the flexibility granted to the quantile curves makes crossing more severe. Postprocessing procedures (e.g. \shortciteNP{dettev08}, \shortciteNP{chernozhukovfg09}) which correct for crossing still suffer from a poor borrowing of information, and
can still lead to wildly variable curves across quantile levels $\tau$. 
\shortciteN{rodrigues2015} proposed a procedure to  postprocess crossing quantiles in a Bayesian framework, however its performance  is still influenced by the initial estimates.  
Recently, several authors have argued that simultaneous estimation offers better estimates, better global efficiency for the estimators have been observed empirically in several studies (\shortciteNP{reichs13}, \shortciteNP{yang2015}, \shortciteNP{fengch2015},  \shortciteNP{rodrigues2016}). 
Indeed simultaneous quantile fitting is a challenging problem that has gained a lot of attention in recent times. A solution to the constrained minimisation problem was proposed by \shortciteN{Bondell2010}, however estimation is again limited to piecewise linear fits. One can also achieve noncrossing by restricting the class of models to the location-scale family (\citeNP{he97}), or by minimising the objective function sequentially while imposing ordering through the parameters (\shortciteNP{Muggeo2013}), but there are naturally limitations when rigid assumptions are imposed. 
More recently, \citeN{yang2015} proposed a Bayesian model for joint estimation of quantile planes using a convenient parameterisation to facilitate the incorporation of monotonicity constraints. Although its focus is linear regression, shrinkage priors for variable selection could potentially be used for fitting splines.

 {Pyramid quantile regression was first presented in \shortciteN{rodrigues2016} as an alternative Bayesian procedure for 
simultaneous quantile fitting. This approach compared favourably with competing methods in empirical studies.}
It relies on the use of several so-called quantile pyramid priors introduced by \citeN{hjortw09} placed at some chosen locations in the covariate space.
By construction this method ensures non-crossing of the different quantile planes within the convex hull of the pyramid locations.  
Although {the spline regression model}   \eqref{eqn:spline} can be estimated under the linear regression umbrella, {the pyramid quantile regression when a large number of basis functions are used can be
problematic. As a matter of fact, the choice of pyramid locations is intricate in this situation since standard convex hull algorithms either cannot cope with high-dimensional  spaces or return too many vertices}. 
This complicates estimation, as crossing of quantile curves will need to be checked at many locations for each parameter update of an MCMC sampling algorithm.

In this paper, we propose a method based on the pyramid quantile regression for simultaneously fitting several penalised spline quantile curves which automatically satisfy the non-crossing constraints.
In order to achieve that, we propose to enclose the data cloud with a polytope whose number of vertices equals the {number of parameters in the regression spline model}, and we derive a general algorithm to find the optimal vertex locations of this convex set, efficiently eliminating all monotonicity constraints. 
As far as we are aware, there are no existing algorithms in the convex optimisation literature which can compute such vertices. Our approach here is also scalable to high dimensions. Next, we develop penalty criterions for estimating flexible quantile curves, and propose an efficient adaptive block-update strategy for MCMC sampling, taking advantage of the proposed convex polytope approach.

The rest of the article is organised as follows. 
Section \ref{subsec:pyrloc} presents the optimal {convex set enclosing the data cloud with the  given number of vertices}. 
Section \ref{sec:quantpyramid} recalls some basics on quantile pyramids introduced by \citeN{hjortw09} and briefly describe their use in \shortciteN{rodrigues2016} in a regression set-up.
In Section \ref{sec:regpyramid} we introduce the penalised quantile splines model and describe the modelling set up. 
Section \ref{sec:sim} presents simulation studies and comparisons with the best alternative approaches. Several real datasets are analysed in section \ref{sec:realex}, and concluding remarks are discussed in the final section.

\section{{Convex set vertices}} \label{subsec:pyrloc}
The convex hull of the predictor cloud has a special role in simultaneous linear quantile regression. 
It is well known that if the conditional quantiles do not cross at the
vertices of a convex set, they do not cross anywhere inside the convex set, see for example \shortciteN{Bondell2010}. 
%The same applies to the case of regression splines, since these can be viewed as multiple linear regression.  
This fact reduces the problem of infinite quantile monotonicity constraints to ensuring monotonicity only at the vertices of the convex set.  \shortciteN{rodrigues2016} used quantile pyramids at selected vertices of the convex hull of the predictor cloud,  non-crossing conditions being checked at the remaining vertices. 
However, in practice, convex hulls of datapoints in dimensions higher than 9 or 10 can be difficult to obtain, and 
when an algorithm is successful at doing so it may return a large number of vertices.  
%it even if they can be obtained efficiently, 
%they may have many vertices 
%, up to the number of observations. 
In these cases, it may be better to work with a larger convex set which 
gives fewer vertices and, consequently, reduced constraints. 
Nevertheless,  some caution needs to be taken when imposing noncrossing constraints on a too large simplex , as it can result in parallel quantile planes (see \citeNP{yang2015}). This point is particularly relevant when covariates are correlated. 

The spline regression model   \eqref{eqn:spline}, viewed as a multiple linear regression model, is an example of such a difficult situation.
%This point is particularly relevant when covariates are correlated, which occurs severely in the current context of splines regression. 
Here the predictor data cloud lies on a one-dimensional curve, say ${\cal X}$, in the high dimensional space $\mathbb{R}^{K+P+1}$ and  the number of vertices of the  convex hull of this predictor data cloud is the entire data set.
To avoid all the abovementioned problems the following proposition gives, for the cubic case $P=3$, the optimal convex set with $K+4$ vertices that contains the curve ${\cal X}$ (see Appendix for a proof).

\begin{proposition}
\label{prop:cross}
Without loss of generality, suppose the covariate $x$ lies in $(0,1)$. 
Consider a cubic polynomial splines \textcolor{black}{basis} with $K$ internal knots denoted by $0< \gamma_1< \gamma_2 < ...<\gamma_{K}< 1$. The minimum {volumn} convex set with $K+4$ vertices that encompasses the curve ${\cal X}$ is a polytope in $\mathbb{R}^{K+3}$ with the following vertices: \\
\begin{equation*}\begin{array}{ll}
x^1 &= (0,0,0,0,0, ..., 0) \\
x^2 &= (\frac{1}{2},0,0,0,0,...,0) \\
x^3 &= (\frac{2}{3}, \frac{1}{3}, 0,0,0, ...,0) \\
x^4 &= (\frac{2+\gamma_1}{3}, \frac{1+2\gamma_1}{3}, \gamma_1, 0, 0,...,0) \\
x^5 &= (\frac{2+\gamma_2}{3}, \frac{1+2\gamma_2}{3}, \gamma_2, (\gamma_2-\gamma_1)(1-\gamma_1)^2,0, ...,0) \\
x^6 &= (\frac{2+\gamma_3}{3}, \frac{1+2\gamma_3}{3}, \gamma_3, (\gamma_3-\gamma_1)(1-\gamma_1)^2, (\gamma_3-\gamma_2)(1-\gamma_2)^2, 0,...,0) \\
x^7 &= (\frac{2+\gamma_4}{3}, \frac{1+2\gamma_4}{3}, \gamma_4, (\gamma_4-\gamma_1)(1-\gamma_1)^2, (\gamma_4-\gamma_2)(1-\gamma_2)^2, (\gamma_4-\gamma_3)(1-\gamma_3)^2,0,...,0) \\
  ... \\
x^{K+3} &= (\frac{2+\gamma_{K}}{3}, \frac{1+2\gamma_{K}}{3}, \gamma_{K}, (\gamma_{K}-\gamma_1)(1-\gamma_1)^2, (\gamma_{K}-\gamma_2)(1-\gamma_2)^2, ...,(\gamma_{K}-\gamma_{K-1})(1-\gamma_{K-1})^2,0) \\
x^{K+4} &= (1,1,1,(1-\gamma_1)^3,(1-\gamma_2)^3,...,(1-\gamma_{K})^3)
\end{array}\end{equation*}
\end{proposition}

\vskip 1cm
Figure \ref{dessins} in the Appendix illustrates the choice of the vertex locations  
for the first elements of the cubic polynomial spline basis, $\{1,x,x^2\}$  and $\{1,x,x^2,x^3\}$. The corresponding convex hull are the shaded regions enclosing, as narrowly as possible,  the curves $\{(x,x^2),x\in (0,1)\}$ and $\{(x,x^2,x^3),x\in (0,1)\}$. The choice of fixed $K+4$ vertices is related to our modelling framework (detailed in the next sections) {and reflects the spline curve degree of freedom}.
%, so that if monotonicity hold at these locations, then linear planes which pass through these points will not cross each other inside the convex hull. 
Proposition \ref{prop:cross} can of course also be used with other modelling setups to ensure non-crossing.

Proposition \ref{prop:cross} provides the vertices with respect to the cubic polynomial spline basis. {For B-splines basis functions, a transformation of basis can be applied to the aforementioned vertices to find the corresponding ones in the new coordinate system (see \shortciteN{book:semipar} and \shortciteN{Spiriti13}). Although the same fit is obtained with both basis, in many applications, the B-spline basis are preferred for computational reasons and will be used in this paper.} More specifically, we replace Equation \eqref{eqn:spline} with cubic B-splines to nonparametrically model the quantile curve,
\begin{eqnarray}
f_\tau(x) = \sum_{j=1}^{K+4} \theta_\tau^j B_j(x) \, ,
\label{bspline}
\end{eqnarray}
where $B_j(x)$, $j=1,...,K+4$, are the B-splines basis functions with $K$ internal knots, and $\theta_{\tau}^j$ denotes the corresponding
$j$th coefficient for the $\tau$th quantile curve. The corresponding vertex locations under the B-splines basis are easily obtained by a simple linear mapping corresponding to the change of basis, see  \shortciteN{book:semipar}. In the next sections, we will describe our modelling approach for the simultaneous penalised quantile spline curves.

\section{Quantile pyramids} \label{sec:quantpyramid}
\textcolor{black}{Here} we provide some background on quantile pyramids, which was introduced by \citeN{hjortw09}. Quantile pyramid is a method for constructing a random probability measure via the quantile function. 
We will use hereafter these priors for  simultaneous inference on quantile spline curves, in the spirit of the linear quantile regression model recently proposed in \shortciteN{rodrigues2016}.
%Recently, \shortciteN{rodrigues2016} proposed linear quantile regression models using these priors for the underlying distribution of the data. Here we present a brief summary of these works for the reader.

Quantile pyramid is a tree generation process with $M$ levels where, at each level $m$, quantiles for fixed probabilities $\tau_{mj}$, $j=1,2, ... ,2^{m-1}$, are randomly generated. 
Consider first that the process starts with $Q_0=0$ and $Q_1=1$, so that the quantile pyramid defines a random distribution on $[0,1]$. Then, the sampling order of the set of given quantile levels $\tau_1<\tau_2<...<\tau_T$ is defined as follows. At $m=1$, we draw a single quantile $Q_{\tau_{11}}$, whose level $\tau_{11}$ is halfway into this set of given levels. Now let ${\tau_{mj}^L}$ and ${\tau_{mj}^R}$ be the closest left and right ancestors of the next quantile level to be drawn $\tau_{mj}$ (ie. antecessors whose quantile levels are adjacent to $\tau_{mj}$). So for the subsequent levels ($m>2$), we draw quantiles $Q_{\tau_{mj}}$, where $\tau_{mj}$ is chosen as the middle level between ${\tau_{mj}^L}$ and ${\tau_{mj}^R}$. If there is an even number of quantile levels to split, for identification purposes, we take the middle value to be the smallest level. The process stops at a finite level $M$, when all quantile levels of interest have been sampled, and the random quantile function is obtained by linear interpolation on the set of quantiles $Q_{\tau_t}$, $t=1,...,T$. %See \shortciteN{rodrigues2016} for further details.

Regarding the random generation process of quantile $Q_{\tau_{mj}}$, this is dictated by 
\begin{equation}
Q_{\tau_{mj}}=Q_{\tau_{mj}^L}(1-V_{mj}) + Q_{\tau_{mj}^R}V_{mj} \, ,
\label{eq:Qm}
\end{equation}
where $V_{mj}$ is a random variable on $(0,1)$ that defines the weights of the averaging process at level $m$. Following \shortciteN{hjortw09} and \shortciteN{rodrigues2016}, we assume that $V_{mj} \sim \text{Beta}(a_{mj},b_{mj})$, with $a_{mj}=2m$ and expected value given by
\begin{equation}
E(V_{mj}) =  \frac{\tau_{mj} - \tau_{mj}^L} {\tau_{mj}^R - \tau_{mj}^L}.
\label{eq:EVgen}
\end{equation}
This describes a quantile process centred on a standard Uniform distribution, denoted thereafter as $Q_\tau^{unif}$, where the variance of the variables $V_{mj}$ decreases with $m$. %and with decreasing prior variance for the quantiles and growing $m$. 

Furthermore, one can centre the quantile process in any distribution with cdf $F$ by applying the following transformation
\begin{equation}
Q_{\tau} = F^{-1}(Q^{unif}_{\tau}) \, .
\label{eq:Npyr}
\end{equation}
Therefore, prior knowledge about the distributional form of the data can be incorporated into the model via $F$. Here we will consider data arising from the reals and use as default centring choice  the Gaussian distribution.%, ie. ${\cal N}(\mu, \sigma)$. 

The simultaneous prior density for the quantiles can be expressed as
\begin{equation}
\begin{array}{ll}
\pi\left( Q_{\tau_1}, ... ,Q_{\tau_T} \right)  = \underset{m=1}{\overset{M}{\prod}}
  \underset{j=1}{\overset{2^{m-1}}{\prod}} \pi_{mj} \left( Q\left(\tau_{mj} \right) \mid Q\left(\tau_{mj}^L\right), Q\left(\tau_{mj}^L\right) \right) \; ,
\end{array}
\label{eq:prior}
\end{equation}
where the conditional densities $\pi_{mj}$ are derived from the transformation of variables given in Equations \eqref{eq:Qm} and \eqref{eq:Npyr}, and considering $V_{mj} \sim \text{Beta}(a_{mj},b_{mj})$. Samples from this quantile process are then piecewise Normal quantile functions. 
For further details, we refer to \citeN{hjortw09}. 
In particular they give conditions for the Bayesian consistency of the  procedure when one uses as prior on the distribution of the data a quantile pyramid with infinite level $M=+\infty$.  

{The pyramid quantile regression model described in \shortciteN{rodrigues2016} uses these pyramid quantile priors  in a regression setting.
In short}, independent pyramid quantile  placed at different locations of the predictor space define the quantile planes. This offers a method for simultaneous inference on the quantile planes, that are by nature non crossing on the convex hull of the pyramid locations. 
{We refer also to this article for a discussion on the posterior consistency of the procedure.}

\section{Quantile Pyramids for Penalised Splines} \label{sec:regpyramid}
The proposed model for simultaneously fitting penalised splines using quantile pyramids is discussed in this section. We consider jointly modelling quantile curves, for a number $T$ of finite quantile levels $\tau= \tau_1< \tau_2 <...< \tau_T$, as a function of a covariate $x\in \mathbb{R}$.

\subsection{Model formulation} \label{subsec:model}
For given quantile levels $\tau$, we consider the use of cubic splines to the model quantile curves. Denote the  $K$ internal knots as $\gamma_1=\gamma_2=\gamma_3=\gamma_4<\gamma_5<...<\gamma_{K+4}<\gamma_{K+5}=\gamma_{K+6}=\gamma_{K+7}=\gamma_{K+8}$, so that $Q_\tau(Y|x)=\sum_{j=1}^{K+4} \theta_\tau^j B_j(x)$, where $B_1, ..., B_{K+4}$ are the B-splines basis functions defined by these knots. 
%Following a similar setup given in \shortciteN{rodrigues2016} for linear regression modelling
For each $\tau$, such linear combinations can be seen as a hyperplane in $\mathbb{R}^{K+4}$, and represented as an affine combination of $K+4$ points $Q_\tau(y|x^p)$, or in short $Q_\tau^p$, located at distinct locations $x^p \in \mathbb{R}^{K+4}$, $p=1, ..., K+4$, that is,
\begin{equation}
Q_{\tau}(Y|x) = \sum_{p=1}^{K+4} Q_\tau^p M_p(x) =  \mathbf{Q}_\tau^p \, \mathbf{M} \; ,
\label{eq:regmodel}
\end{equation}
where the design matrix $\mathbf{M}$ can be obtained from $\mathbf{M} = (\mathbf{Q}_\tau^p)^{-1} \, \mathbf{Q}_{\tau}(Y|x)$, for arbitrary points $\mathbf{Q}_\tau^p$ and the corresponding quantiles $\mathbf{Q}_{\tau}(Y|x_i)$, $i=1, ..., n$. 

Since model \eqref{eq:regmodel} is parameterised in terms of the quantiles $Q_{\tau}^p$, we consider using $K+4$ separate quantile pyramids to represent the prior quantile processes at each $x^p$ location.  We use as the pyramid locations the vertices of the convex hull given in Proposition \ref{prop:cross}.
This ensures that
%, and since the quantile pyramids aremonotone at each location, then 
the hyperplanes that passes through these points at varying quantile levels $\tau$ will not cross inside the convex hull, thereby producing
non-crossing regression splines without any additional need to check for non-crossing during computation.

%can be chosen such that the fitted
%quantile curves will not cross,  we discuss the optimal choice for the pyramid locations in section \ref{subsec:pyrloc}.

Assuming pyramid priors centred on the Normal distribution, the corresponding likelihood is formulated as a piecewise Normal density
\begin{equation}
f(y |{x})= \sum_{t=1}^{T} {(\tau_t-\tau_{t-1})\frac{\phi(y; \mu_{x},\sigma_{ x}^2)}{ \Phi \left (\frac{Q_{\tau_t}(Y|{x})-\mu_{{x}}}{\sigma_{{x}}}\right )-\Phi \left(\frac{Q_{\tau_{t-1}}(Y|{x})-\mu_{{x}}}{\sigma_{{ x}}}\right)}I_{(Q_{\tau_t}(Y|{x}), Q_{\tau_{t-1}}(Y|{x})]}(y)},
\label{eqnLik}
\end{equation}
where $I_{(q_1, q_2]}(y)$ is 1 if $y \in (q_1, q_2 ]$ and zero otherwise, and {$\phi(\cdot; \mu_x,\sigma_x^2)$} denotes the density function of the centring Normal distribution {${\cal N}(\mu_x,\sigma_x^2)$ , where the centring mean $\mu_x$ and the centring
standard deviation $\sigma_x$  can vary with $x$. We describe how to
specify the centring parameters and penalisation in the next sections.}

\subsection{Penalised centring mean} \label{subsec:pyrmean}
%\subsection{Penalisation}  \label{subsec:pyrmean}
% it is important to let the parameters of the centring distribution vary with the explanatory variable, 
For nonlinear curve fitting, we consider allowing the centring mean $\mu_x$ to also vary flexibly and smoothly with $x$. To achieve this,
we assume that the mean curve of the centring distribution is a cubic B-spline with the same set of knots $\gamma$, ie. $\mu_{\bf x}= \sum_{j=1}^{K+4} \eta_j B_j(x)=\boldsymbol{\eta} \mathbf{B}$, as for the quantiles. It is well known that the number and location of knots in spline based regression models play an important role in obtaining a good fit. Here we use a large number of knots $K$, and penalise the centring mean curve to obtain smoothness. Note that by constraining the mean and variance parameters of the centring distribution to vary smoothly across $x$, we obtain smoothness for the entire centring distributions (ie. all centring quantiles).

{In order to avoid rank deficient covariance matrix and facilitate the incorporation of more complex models, we use the O'Sullivan penalised splines. See  \citeN{Wand2016} for a discussion on the differences between P-splines and smoothing splines.  Their mixed model formulation 
%from \citeN{Wand2016} 
is considered here,}
\begin{subequations}
\begin{align}
\mu_{x} &= \mathbf{X} \boldsymbol{\beta} + \mathbf{Zu} \label{eq:mucentera} \; ,  \\
\mathbf{u} &\sim {\cal N}(\mathbf{0}, \sigma^2_u \mathbf{I})\label{eq:mucenterb}  \; , 
\end{align}
\end{subequations}
with design matrices $\mathbf{X}_{(N\times2)}=[1,x_i]_{1\leq i \leq n}$ and $\mathbf{Z}_{(N\times (K+2))}=\mathbf{B}\mathbf{U}_Z diag(\mathbf{d}_Z^{-1/2})$, for the fixed and random effects, respectively. In order to obtain $\mathbf{U_Z}$ and $\mathbf{d}_Z$, consider the $(K+4)\times(K+4)$ penalty matrix $\boldsymbol{\Omega}$, whose entries are $\boldsymbol{\Omega}_{kk'} = \int_a^b B_k''(x) B_{k'}''(x) dx$, and take its spectral decomposition $\boldsymbol{\Omega} = \mathbf{U} \, diag(\mathbf{d}) \, \mathbf{U}^T$. Here $\mathbf{U}^T\mathbf{U}=\mathbf{I}$, and $\mathbf{d}_Z$ is the $(K+2)$ sub-vector of $\mathbf{d}$ containing its positive entries, whereas $\mathbf{U}_Z$ is the $(K+4)\times(K+2)$ sub-matrix of $\mathbf{U}$ with columns corresponding to the positive entries of $\mathbf{d}$. For further details, including codes, see \citeN{Wand2016}. {Following their suggestions, we adopt throughout the paper standard non-informative priors for the additional parameters $\boldsymbol{\beta}$ and $\sigma^2_u$, $\beta_0 \sim {\cal N}(0, 10^8)$, $\beta_1 \sim {\cal N}(0, 10^8)$ and $\sigma^2_u \sim IG(0.01, 0.01)$, but results are not sensitive to these choices.}

\subsection{Centring standard deviation} \label{subsec:pyrvar}
A simpler cubic B-splines is considered for modelling the standard deviation of the centring distribution, as generally the variability function has less fluctuation and complexity. Therefore, we consider $\sigma_x = \sum_{j=1}^{R+4} \nu_j B_j(x)$, where $R$ is a reduced number of interior knots. Using the hyperplane parameterisation we have
\begin{equation}
\sigma_x = \sum_{p=1}^{R+4} \sigma^p N_p(x) = \boldsymbol{\sigma}^p \mathbf{N} \; ,
\label{eq:sdcenter}
\end{equation}
where $\mathbf{N}$ is the corresponding design matrix and $\boldsymbol{\sigma}^p$ is a vector containing the standard deviations at $(R+4)$ pyramids. Independent uniform priors were adopted for the standard deviations, ${\sigma}^p \sim U(0.01, 10^6)$, and $R=3$ internal knots were used for all simulations and applications, providing sufficient flexibility for the variability curves.

\subsection{{Model fitting}} \label{subsec:mcmc}
Based on these prior specifications, the proposed model can be summarized as follows
\begin{subequations}
\begin{align}
Y|x \sim  \; pw{\cal N} &(\mathbf{Q}_{\tau_t, t=1:T}(Y|x),  \mu_x,  \sigma_x) \label{pwn} \textcolor{black}{\qquad}\\
Q_{\tau_t}(Y|x) =& \,\mathbf{Q}_{\tau_t}^p \, \mathbf{M}, \; \forall t=1,...,T \label{qmodel} \\
\mu_{x} = \mathbf{X} \boldsymbol{\beta} + \mathbf{Zu}  \; &, \; 
\mathbf{u} \sim {\cal N}(\mathbf{0}, \sigma^2_u \mathbf{I}) \label{mumodel} \\
\sigma_x &= \boldsymbol{\sigma}^p \mathbf{N} \label{sigmamodel}
\end{align}
\end{subequations}
where \textcolor{black}{$pw{\cal N}(\mathbf{Q}_{\tau_t, t=1:T}(Y|x), \mu_x,\sigma_x)$ is the distribution with piecewise Normal density \eqref{eqnLik}, whose parameters $\mu_x$ and $\sigma_x$} are functions of covariate $x$. More specifically, quantiles $Q_{\tau_t}(Y|x)$  are cubic B-splines with $K$ knots \eqref{qmodel}, mean function $\mu_{x}$ is a penalised cubic B-spline \eqref{mumodel}, and $\sigma_x$ is a cubic B-spline with reduced $R$ knots \eqref{sigmamodel}. And with respect to the prior choices, quantile pyramid is considered for $Q_{\tau_t}(Y|x)$, whereas for the other parameters we use $\beta_0,\beta_1 \sim {\cal N}(0, 10^8)$, $\sigma^2_u \sim IG(0.01, 0.01)$, ${\sigma}^p \sim U(0.01, 10^6)$.

% \begin{subequations}
% \begin{align*}
% Q_\tau^p &\sim PyrP (\boldsymbol{\beta}, \mathbf{u}, \sigma_u^2,  \boldsymbol{\sigma}^p) \\
% \beta_0,\beta_1 &\sim {\cal N}(0, 10^8) \\
% \sigma^2_u &\sim IG(0.01, 0.01) \\
% {\sigma}^p &\sim U(0.01, 10^6)
% \end{align*}
% \end{subequations}

Model inference is based on running Markov Chain Monte Carlo over the parameter vector $(\mathbf{Q}_{\tau_t}^p, \boldsymbol{\beta}, \mathbf{u}, \sigma_u^2,  \boldsymbol{\sigma}^p)$, whose elements have dimensions $(K+4)\times T, 2 , (K+2) , 1, R+4$, respectively. %Considering that the parameters are very correlated to each other, 
{When the number of covariates is large, parameters become highly correlated and more advanced MCMC techniques become necessary. Placing quantile pyramids
at the convex hull vertices produces non-crossing planes by construction, this also means that constructing an adaptive MCMC algorithm in which we make use of parameter
correlation structure is now feasible. It turns out that learning the correlation is crucial to MCMC performance. We propose to do this is in a two stage procedure.
}
%{Due to the nature of high correlations between some of the parameters, an adaptive MCMC algorithm is proposed in order to explore the posterior distribution more efficiently.} 
In a first stage, parameters are updated one at a time using the Metropolised Gibbs sampling. Proposal distributions for quantiles $\mathbf{Q}_{\tau_t}^p$ are Uniform, and for the remaining parameters are Gaussian, all centred on the current value of the chain.  A Robbins-Monro search scheme algorithm detailed in  \shortciteN{garthwaitefs11} is used to automatically tune the scaling parameters of these proposal distributions in order to achieve the optimal acceptance rate of 0.44 (\shortciteNP{roberts+r01}).%, details of the algorithm can be found in \shortciteN{garthwaitefs11}. 
The covariance structure among model parameters is estimated based on the first stage sampling, and hierarchical clustering is performed to create blocks based on this posterior correlation (using Blocks function from R-package LaplacesDemon, \shortciteNP{LaplacesDemon}). Then, in a second stage, block-wise Metropolis-Hastings is performed using a multivariate Normal proposal distribution, with previously estimated posterior correlation matrix. Again, we consider the algorithm of \shortciteN{garthwaitefs11}  to tune the scaling in order to achieve the multivariate optimal acceptance rate of 0.23 (\shortciteNP{roberts1997}). Parameter estimates are then calculated considering the posterior mean of the chain.

\section{Simulated examples} \label{sec:sim}
In this section we examine the performances of the proposed method {\it via} a simulation study. We compare its results  with the ones obtained by recent alternative approches with available codes.
%In this section, simulation studies are presented in order to explore finite sample properties of the proposed estimates and compare it with the best alternative approaches.  We restrict the comparisons to methods with codes available. 
This includes constrained B-splines smoothing (COBs), which provides individual quantile fittings using quadratic splines with $L_\infty$ penalty,  available in the  R package cobs, see \shortciteN{COBSpackage} and  \shortciteN{Pin07}. 
This includes also the simultaneous quantile curve estimation available from the {\sf gcqr} function of the quantregGrowth R-package (\shortciteNP{Muggeo2013}), which provides cubic p-splines with $L_2$ penalty. Here standard errors of the estimates under this method were calculated using 100.000 bootstrap samples. 

In the Bayesian framework, \citeN{yang2015} recently proposed a model for joint estimation of quantile planes (QRJ from R-package qrjoint, \citeNP{qrjoint}). QRJ estimates were obtained here using $40.000$ MCMC samples, thinning every $10$ samples and discarding the initial $20\%$ of the samples as burn-in, also $\tau$ increment was set to $0.001$. Although QRJ primary focus is linear regression, the authors point out that their modelling platform is broad and discuss the use of shrinkage priors for variable selection. Therefore, we attempted here to use B-spline transform of the covariate variable as linear predictor in the regression model, and applied the suggested shrinkage priors (horseshoe prior for $\gamma_0$ and $\gamma$, and a spike-slab mixture of gamma for $\kappa$). 

Estimates from the method presented in this paper, pyramid quantile penalised splines (PQPS), are obtained from an adaptive Markov chain Monte Carlo sampler. 
%As a result of the high correlation among model parameters, which is characteristic of simultaneous quantile regression, an adaptive MCMC is designed to optimise the algorithm performance. 
The underlying covariance structure is estimated considering $60.000$ MCMC draws and burn-in of $10.000$. Thereafter blockwise sampling is performed with $200.000$ MCMC draws, thinning every $10$ samples, and $10.000$ burn-in. For all aforementioned approaches we consider fitting a B-spline curve with $20$ knots spread evenly across covariate values. 

Codes for fitting noncrossing splines regression quantiles (NCRQ) by solving a constrained minimisation problem (\shortciteNP{Bondell2010}) are not available, so we included here the nonparametric simulation designs proposed by them to have their results for comparison. NCRQ fits linear splines with the total variation penalty, and adopted in the simulation study B-splines with 25 knots. Therefore, following \shortciteN{Bondell2010}, a sample of $n=100$ observations is drawn from a heteroscedastic error model $y_i = f(x_i) + g(x_i)\epsilon_i$, with $x_i$ sampled from the $U(0,1)$ distribution and  $\epsilon_i$ sampled from the $N(0,1)$ distribution, and the following choices of mean and covariance functions:

\begin{description}
 \item [Design 1.] $f(x)=0.5+2x+\text{sin}(2\pi x-0.5)$, $g(x)=1$;
 \item [Design 2.] $f(x)=3x$, $g(x)=0.5+2x+\text{sin}(2\pi x-0.5)$;
\end{description}

To compare the methods, $200$ data sets were simulated and empirical root mean integrated squared error, $\text{RMISE}=\sqrt{1/n\sum_{i=1}^n{\{Q(\tau|x_i) - \widehat{Q}(\tau|x_i)\}^2}}$, and $95\%$ (frequentist) coverage probabilities were computed for quantile levels $\tau=0.5,0.7,0.9,0.95,0.99$. Results for Designs 1 and 2 are presented in Tables \ref{tab:des1} and \ref{tab:des2}, respectively. Note that results for NCQR quantile estimates at $\tau=0.5,0.7,0.9$ are borrowed from \shortciteN{Bondell2010}, and coverages were not reported. 

\begin{table}[!ht] \centering 
  \caption{RMISE ($\times 100$) and $95$\% coverage probabilities for Design 1} 
  \label{tab:des1} 
\begin{tabular}{@{\extracolsep{5pt}} D{.}{.}{-2} D{.}{.}{-2} D{.}{.}{-2} D{.}{.}{-2} D{.}{.}{-2} D{.}{.}{-2} } 
\\[-1.8ex]\hline 
\hline \\[-1.8ex] 
\multicolumn{1}{c}{} & \multicolumn{1}{c}{$0.5$} & \multicolumn{1}{c}{$0.7$} & \multicolumn{1}{c}{$0.9$} & \multicolumn{1}{c}{$0.95$} & \multicolumn{1}{c}{$0.99$} \\ 
\hline \\[-1.4ex]
\multicolumn{1}{l}{\textit{RMISE}} &  &  &  &  &    \\[0.2ex]
\multicolumn{1}{l}{\hspace{0.3cm} PQPS} & 24.5 & 24.9 & 32.4 & 37.2 & 46.6  \\ 
\multicolumn{1}{l}{\hspace{0.3cm} COBs} & 26.9 & 29.0 & 41.0 & 53.3 & 93.3  \\ 
\multicolumn{1}{l}{\hspace{0.3cm} GCQR} & 28.6 & 29.7 & 36.2 & 42.5 & 62.1 \\ 
\multicolumn{1}{l}{\hspace{0.3cm} QRJ} & 36.9 & 37.2 & 39.0 & 40.2 & 47.7  \\
\multicolumn{1}{l}{\hspace{0.3cm} NCQR} & 25.7 & 25.9 & 31.8 &  \multicolumn{1}{c}{-} & \multicolumn{1}{c}{-} \\ [1ex]
\multicolumn{1}{l}{\textit{Coverage}} &  &  &  &  &    \\[0.2ex]
\multicolumn{1}{l}{\hspace{0.3cm} PQPS} & 0.95 & 0.95 & 0.95 & 0.95 & 0.95  \\ 
\multicolumn{1}{l}{\hspace{0.3cm} COBs} & 0.77 & 0.76 & 0.58 & 0.31 & 0.00 \\ 
\multicolumn{1}{l}{\hspace{0.3cm} GCQR} & 0.95 & 0.95 & 0.92 & 0.82 & 0.49  \\ 
\multicolumn{1}{l}{\hspace{0.3cm} QRJ} & 0.94 & 0.95 & 0.96 & 0.97 & 0.97  \\
\multicolumn{1}{l}{\hspace{0.3cm} NCQR} & \multicolumn{1}{c}{-} & \multicolumn{1}{c}{-} & \multicolumn{1}{c}{-} & \multicolumn{1}{c}{-} & \multicolumn{1}{c}{-} \\
\hline \\[-1.8ex] 
\end{tabular} 
\end{table}

\begin{table}[!ht] \centering 
  \caption{RMISE ($\times 100$) and $95$\% coverage probabilities for Design 2} 
  \label{tab:des2} 
\begin{tabular}{@{\extracolsep{5pt}} D{.}{.}{-2} D{.}{.}{-2} D{.}{.}{-2} D{.}{.}{-2} D{.}{.}{-2} D{.}{.}{-2} } 
\\[-1.8ex]\hline 
\hline \\[-1.8ex] 
\multicolumn{1}{c}{} & \multicolumn{1}{c}{$0.5$} & \multicolumn{1}{c}{$0.7$} & \multicolumn{1}{c}{$0.9$} & \multicolumn{1}{c}{$0.95$} & \multicolumn{1}{c}{$0.99$} \\ 
\hline \\[-1.4ex]
\multicolumn{1}{l}{\textit{RMISE}} &  &  &  &  &    \\[0.2ex]
\multicolumn{1}{l}{\hspace{0.3cm} PQPS} & 18.5 & 23.0 & 41.6 & 51.4 & 69.3  \\ 
\multicolumn{1}{l}{\hspace{0.3cm} COBs} & 26.7 & 35.9 & 58.2 & 76.2 & 140.4  \\ 
\multicolumn{1}{l}{\hspace{0.3cm} GCQR} & 41.6 & 45.9 & 57.0 & 67.0 & 105.1  \\ 
\multicolumn{1}{l}{\hspace{0.3cm} QRJ} & 55.9 & 64.5 & 89.5 & 105.2 & 146.3  \\
\multicolumn{1}{l}{\hspace{0.3cm} NCQR} & 26.4 & 32.2 & 48.5 & \multicolumn{1}{c}{-} & \multicolumn{1}{c}{-} \\ [1ex]
\multicolumn{1}{l}{\textit{Coverage}} &  &  &  &  &    \\[0.2ex]
\multicolumn{1}{l}{\hspace{0.3cm} PQPS} & 0.97 & 0.97 & 0.94 & 0.92 & 0.90  \\ 
\multicolumn{1}{l}{\hspace{0.3cm} COBs} & 0.85 & 0.77 & 0.62 & 0.42 & 0.00  \\ 
\multicolumn{1}{l}{\hspace{0.3cm} GCQR} & 0.96 & 0.95 & 0.87 & 0.77 & 0.49  \\ 
\multicolumn{1}{l}{\hspace{0.3cm} QRJ} & 0.91 & 0.91 & 0.87 & 0.85 & 0.86 \\
\multicolumn{1}{l}{\hspace{0.3cm} NCQR} & \multicolumn{1}{c}{-} & \multicolumn{1}{c}{-} & \multicolumn{1}{c}{-} & \multicolumn{1}{c}{-} & \multicolumn{1}{c}{-} \\
\hline \\[-1.8ex] 
\end{tabular} 
\end{table}

\begin{figure}[!ht]
    \centering
    \footnotesize
    \includegraphics[width=4cm,height=4cm,angle=-90]{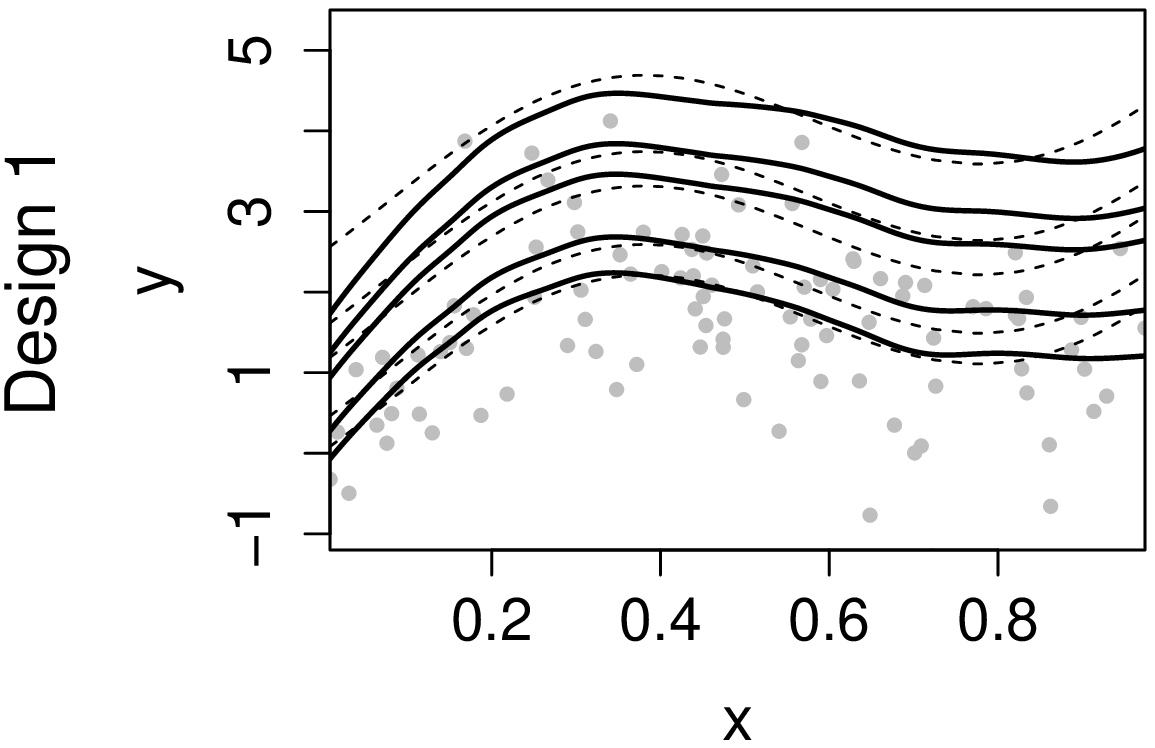}
    \includegraphics[width=4cm,height=4cm,angle=-90]{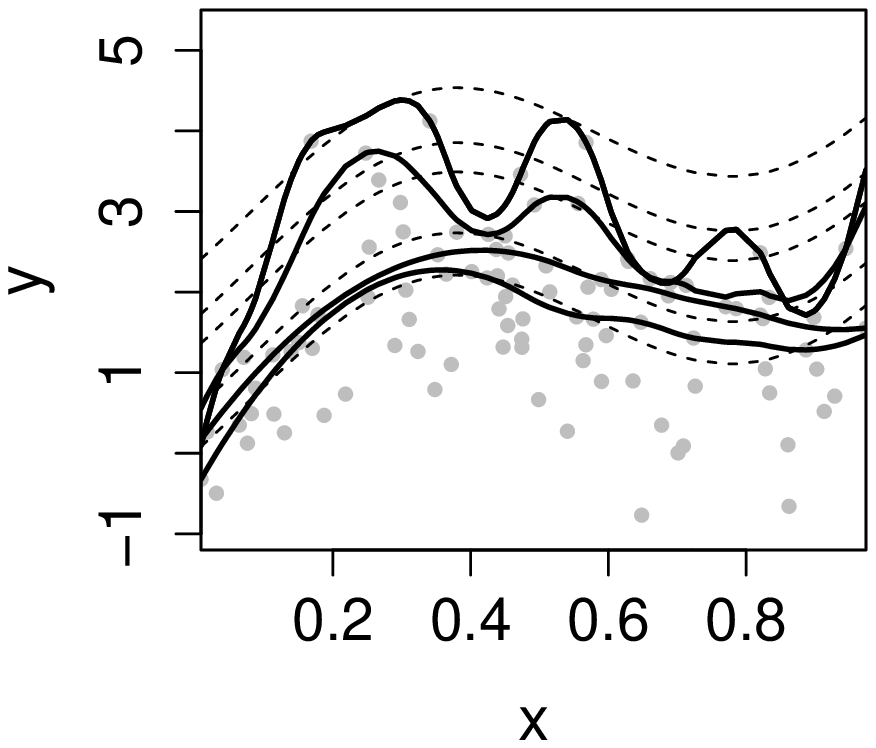}
    \includegraphics[width=4cm,height=4cm,angle=-90]{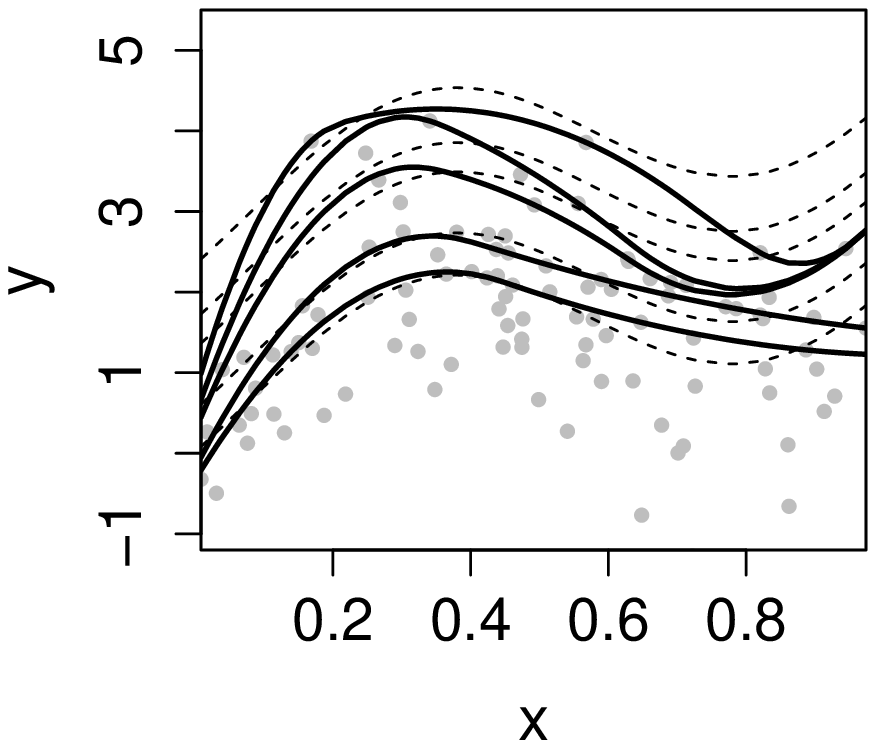}
    \includegraphics[width=4cm,height=4cm,angle=-90]{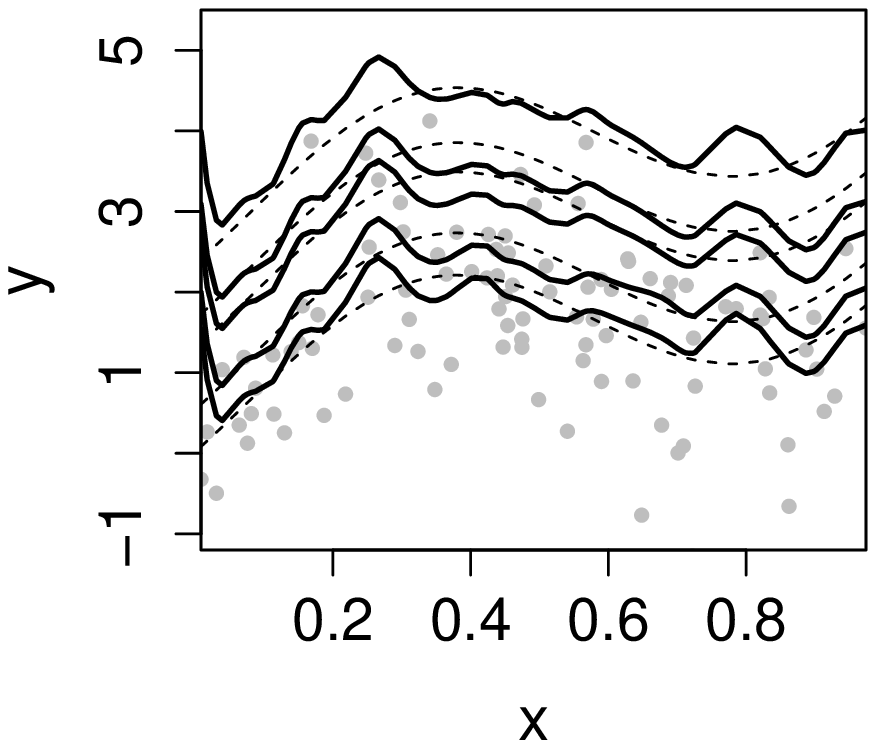} \\ 
    \vspace{-0.5cm}
    \stackunder[5pt]{\includegraphics[width=4cm,height=4cm,angle=-90]{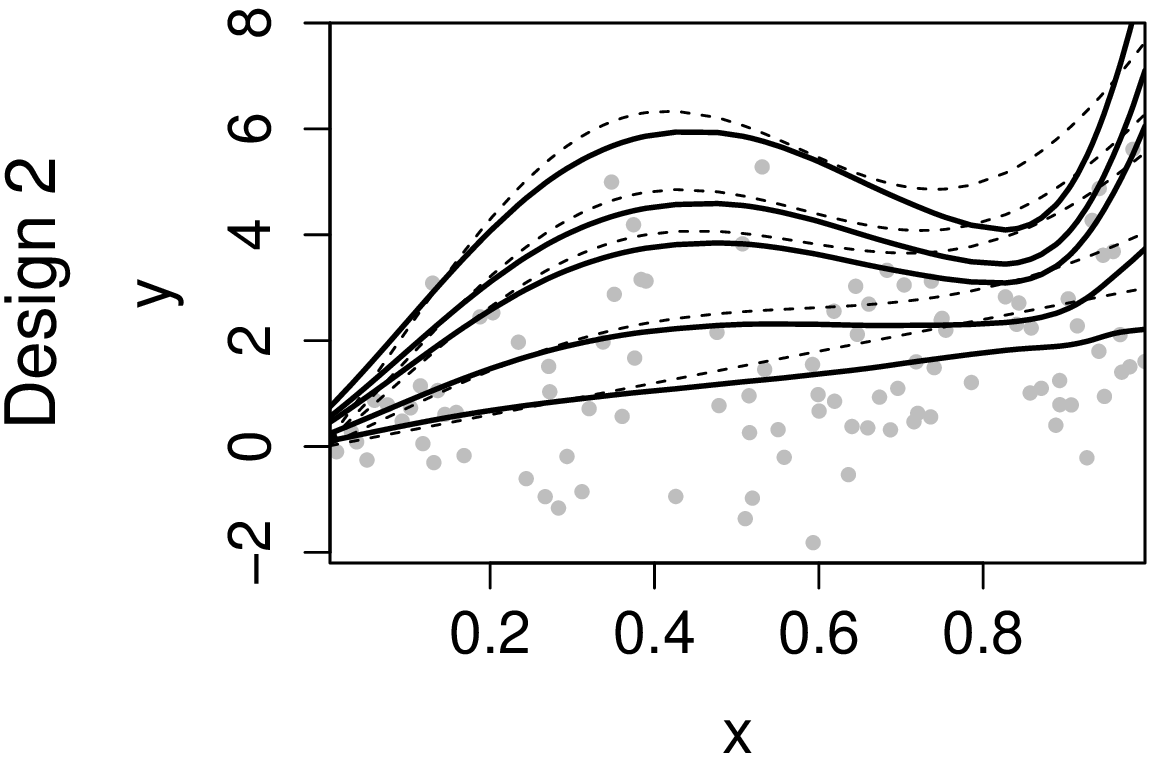}}{\hspace{0.3cm} PQPS}
    \stackunder[5pt]{\includegraphics[width=4cm,height=4cm,angle=-90]{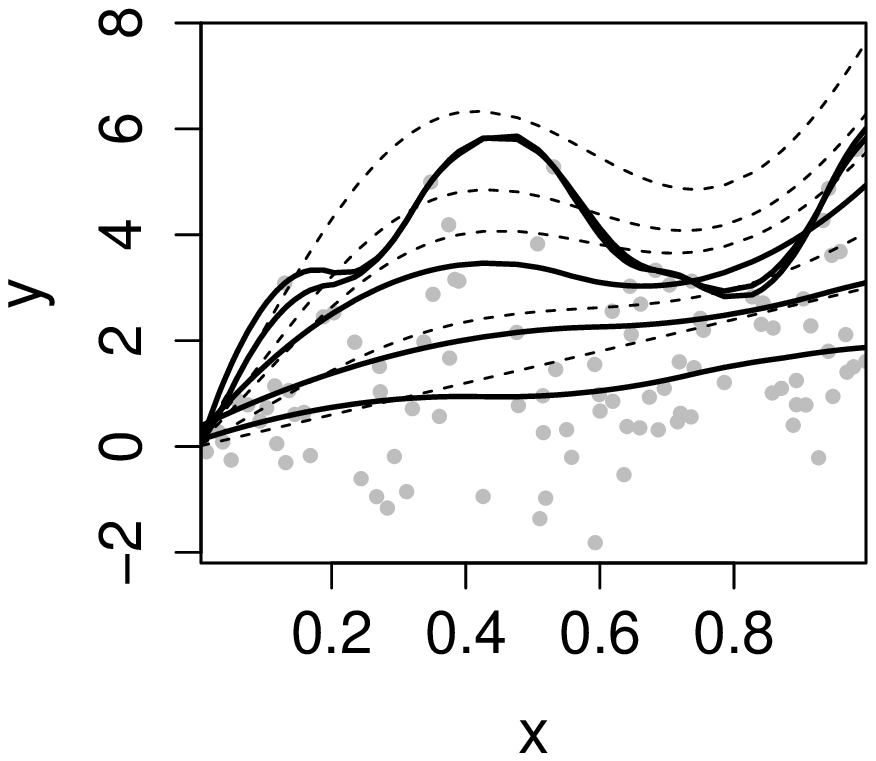}}{\hspace{0.3cm} COBs}
    \stackunder[5pt]{\includegraphics[width=4cm,height=4cm,angle=-90]{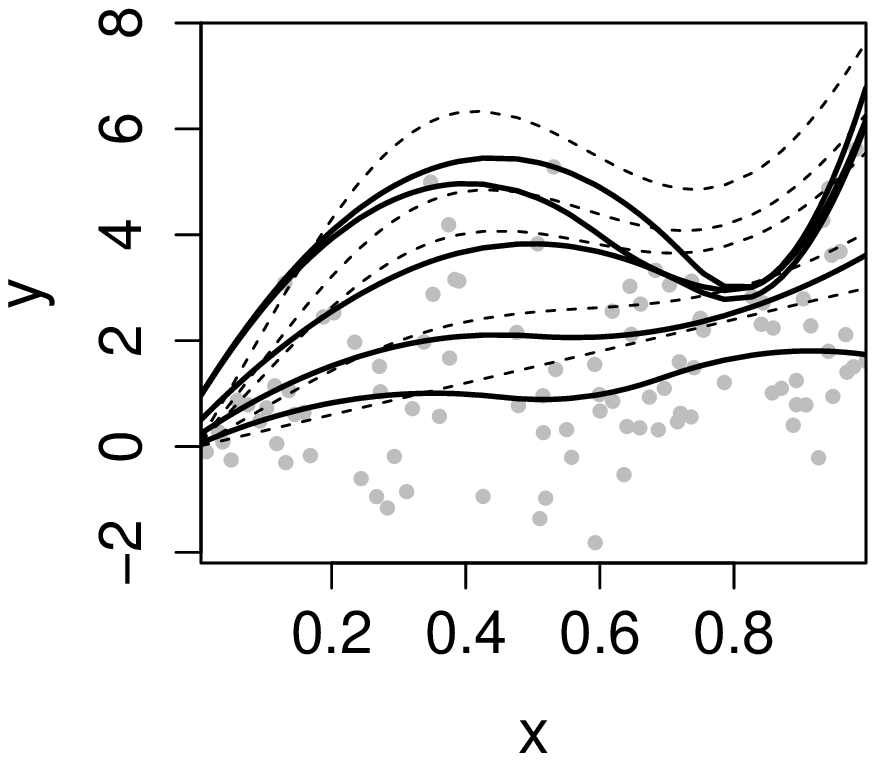}}{\hspace{0.3cm} GCQR}
    \stackunder[5pt]{\includegraphics[width=4cm,height=4cm,angle=-90]{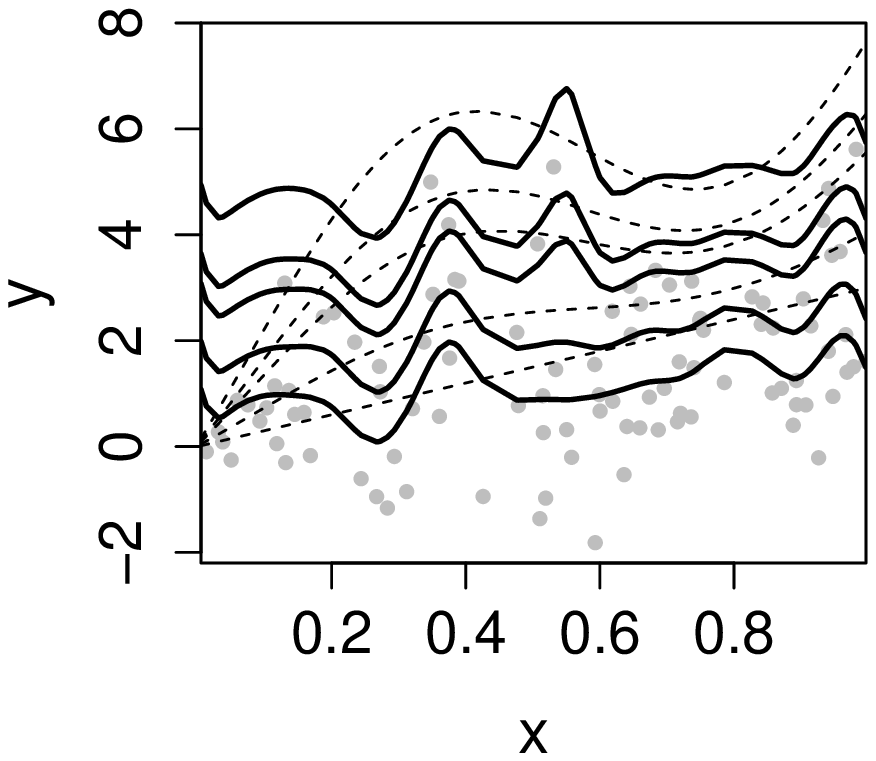}}{\hspace{0.3cm} QRJ}
    
    \caption{Estimated quantile curves at $\tau=0.5,0.7,0.9,0.95,0.99$ for one sample from Design 1 (first row) and Design 2 (second row). Dashed lines are true quantile functions, and solid lines are quantile estimates from pyramid quantile penalised spline (PQPS), constrained B-splines (COBs), growth chart regression quantiles (GCRQ) and joint estimation of linear quantile planes (QRJ).}
    %and RQ (bottom row, Koenker (2005)) at $\tau=0.25,0.5,0.75$. Right column zoom in the initial interval (2 to 12ms).
    \label{fig:des12}
\end{figure}

Design 1 presents a simple scenario in which the variance function is constant, so quantile curves are simply parallel. From Table \ref{tab:des1}, PQPS performs similarly to NCQR in terms of RMISE, both considerably outperforming the other methods. Although coverage probabilities are not reported in \shortciteN{Bondell2010} for NCQR, as shown in \shortciteN{rodrigues2016}, for linear regression this estimate has coverages well below nominal level for quantiles at the tails, and similar behaviour is expected here as inference is also based on asymptotic results. Coverage probabilities for frequentist methods COBs and GCQR are also unsatisfactory for the same reason. QRJ presents nice coverage probabilities, although this current implementation evidently overfits the data, as shown in Figure \ref{fig:des12}, which illustrates estimated quantile curves for one sample from Designs 1 and 2. {We note that QRJ was not developed specifically for nonparametric regression, 
%Naturally, model fit can be compromised and overfitting expected, as there should be more suitable prior choices for smoothing spline fitting than those provided in the R-package. And 
and although we tried different hyperparameters choices, no significant changes in the final fits were observed. Therefore, QRJ results throughout are preliminary and should be interpreted accordingly.}

On the other hand, Design 2 brings an elaborate variance function and very distinct quantile curves for each $\tau$ (see Figure \ref{fig:des12}). For this scenario, Table \ref{tab:des2} shows that PQPS has considerably smaller RMISE than all other methods across all quantile levels, which demonstrates the flexibility of the proposed approach to fit complex quantile functions. In addition, PQPS coverage probabilities are also better than its competitors, being closer to the $95\%$ nominal level. 

For both simulation designs, Figure \ref{fig:des12} shows severe crossings for individual quantile fitting curves from COBs, furthermore its quantile curves are clearly dissociated from one another as no borrowing information is considered. For GCQR, although quantile curves do not cross, curves are also unrelated. Indeed, when quantiles are fitted separately (or merely considering monotonicity constraints), one is discarding valuable information, which is particularly troublesome when interest also lies at the tails of the distribution, where data is already scarce. From Figure \ref{fig:des12}, we also note that the shrinkage priors used in QRJ for variable selection were incapable of discarding irrelevant covariates, so overfitting occurs. 
%Although the framework proposed by \citeN{yang2015} has potential, more effort is necessary in order to properly use it for spline fitting \textcolor{magenta}{[???]}.

\begin{figure}[!ht]
    \centering
    \footnotesize
    \includegraphics[width=4cm,height=4cm,angle=-90]{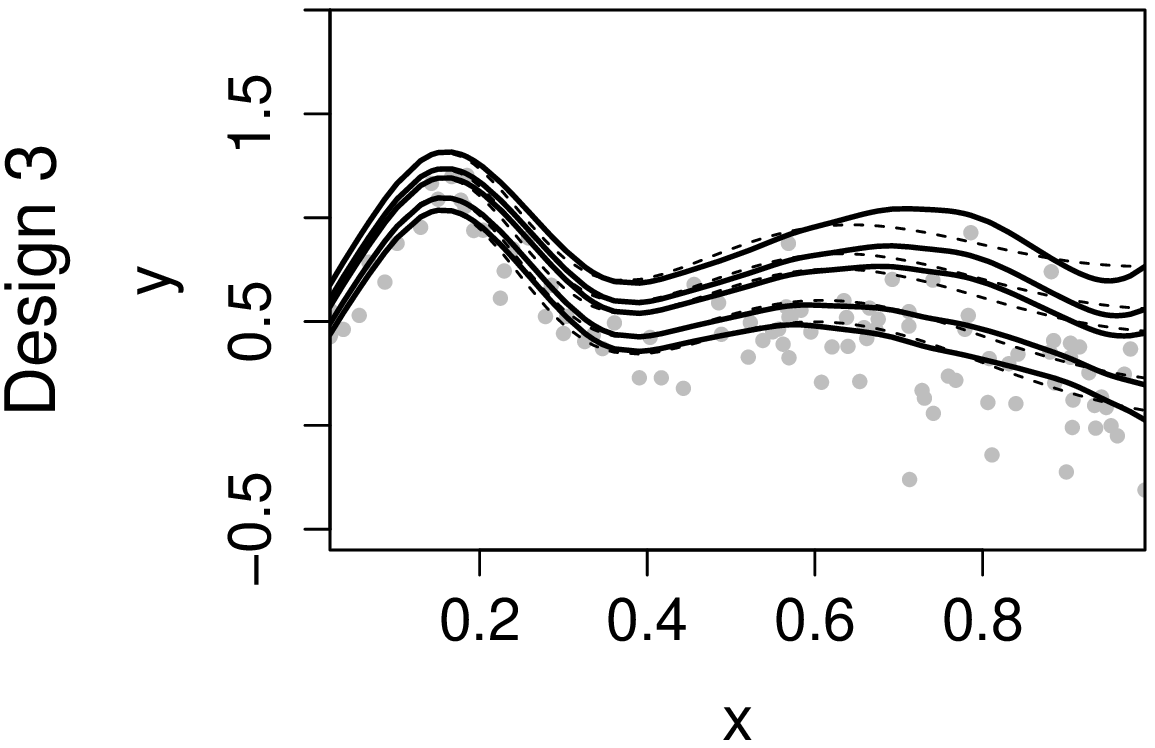} %\textcolor{black}
    \includegraphics[width=4cm,height=4cm,angle=-90]{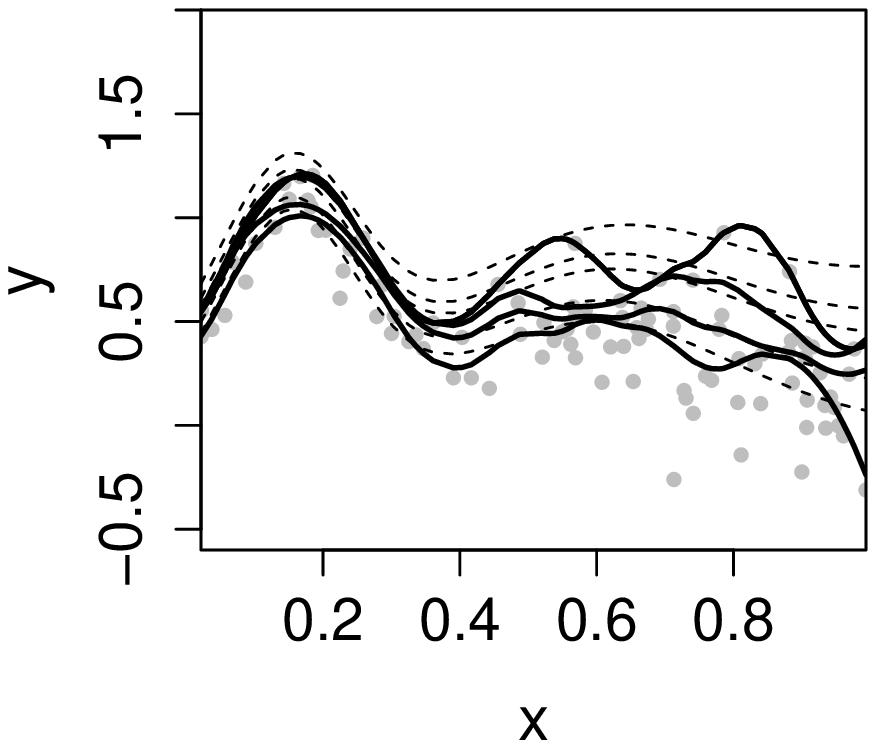}
    \includegraphics[width=4cm,height=4cm,angle=-90]{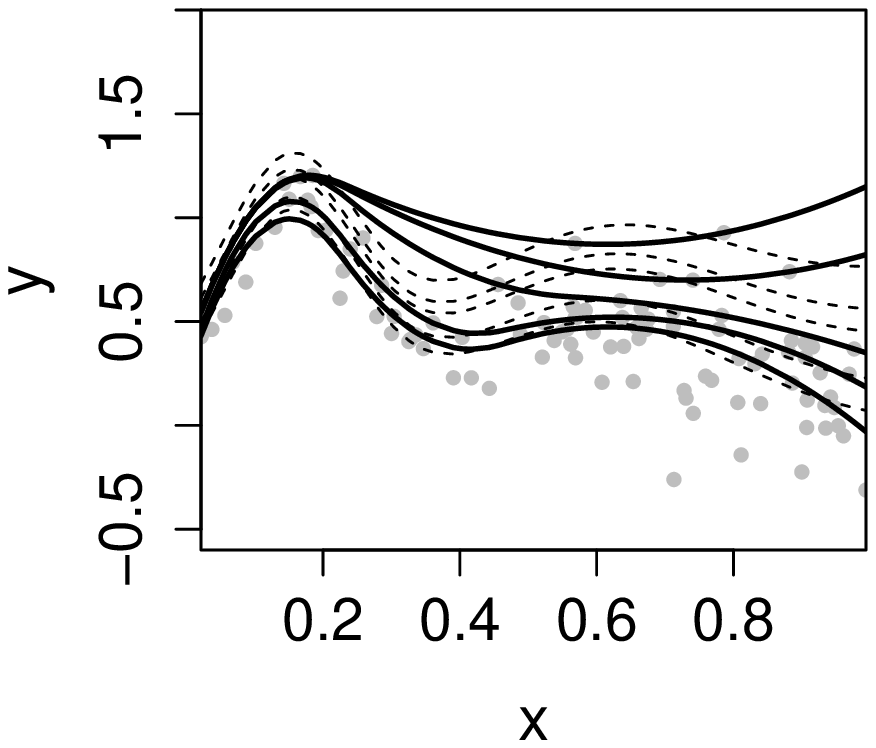}
    \includegraphics[width=4cm,height=4cm,angle=-90]{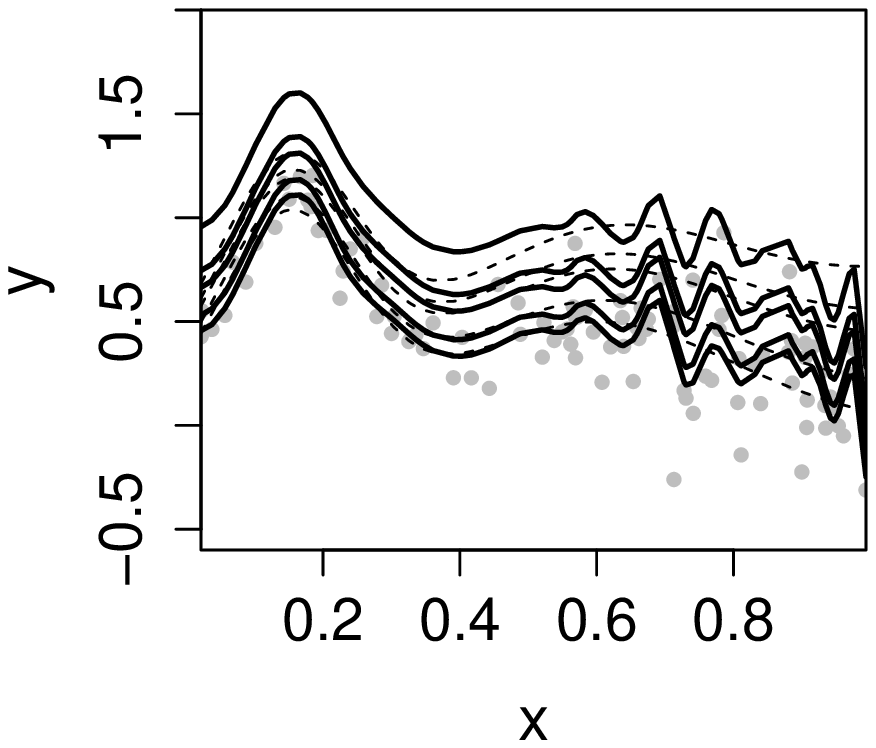} \\ 
    \vspace{-0.5cm}
    \stackunder[5pt]{\includegraphics[width=4cm,height=4cm,angle=-90]{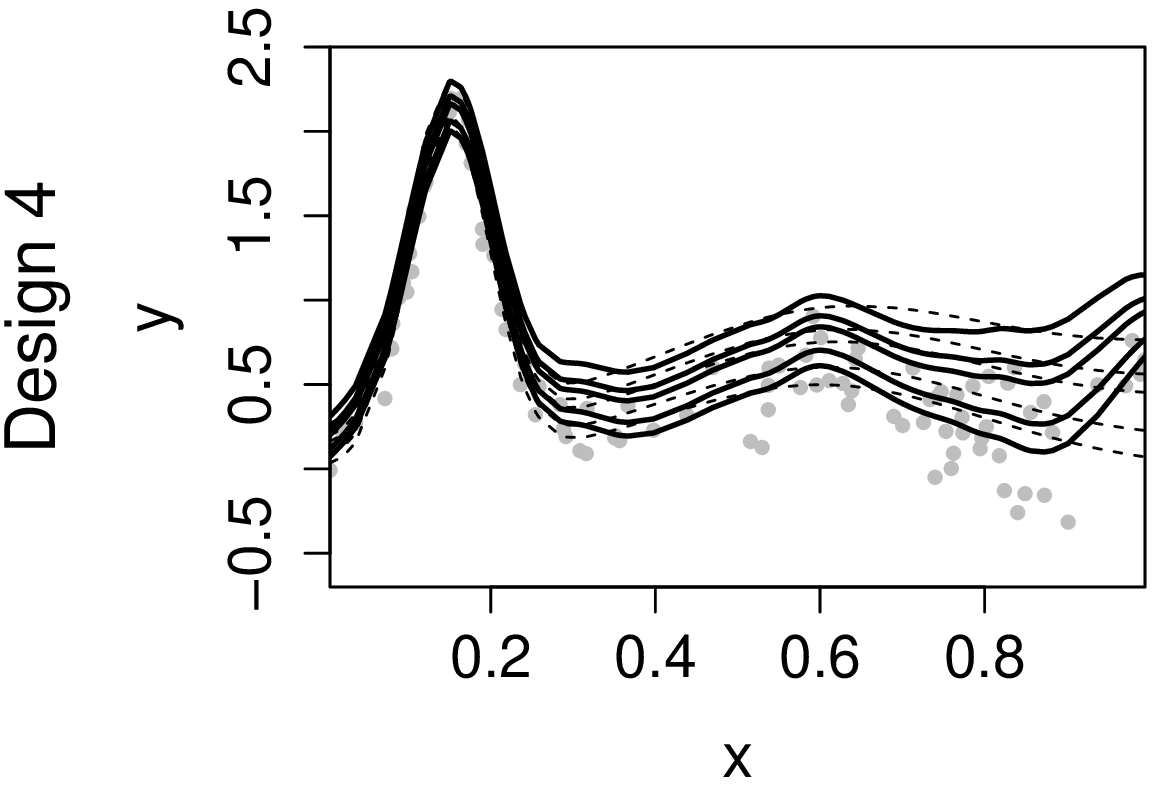}}{\hspace{0.3cm} PQPS}
    \stackunder[5pt]{\includegraphics[width=4cm,height=4cm,angle=-90]{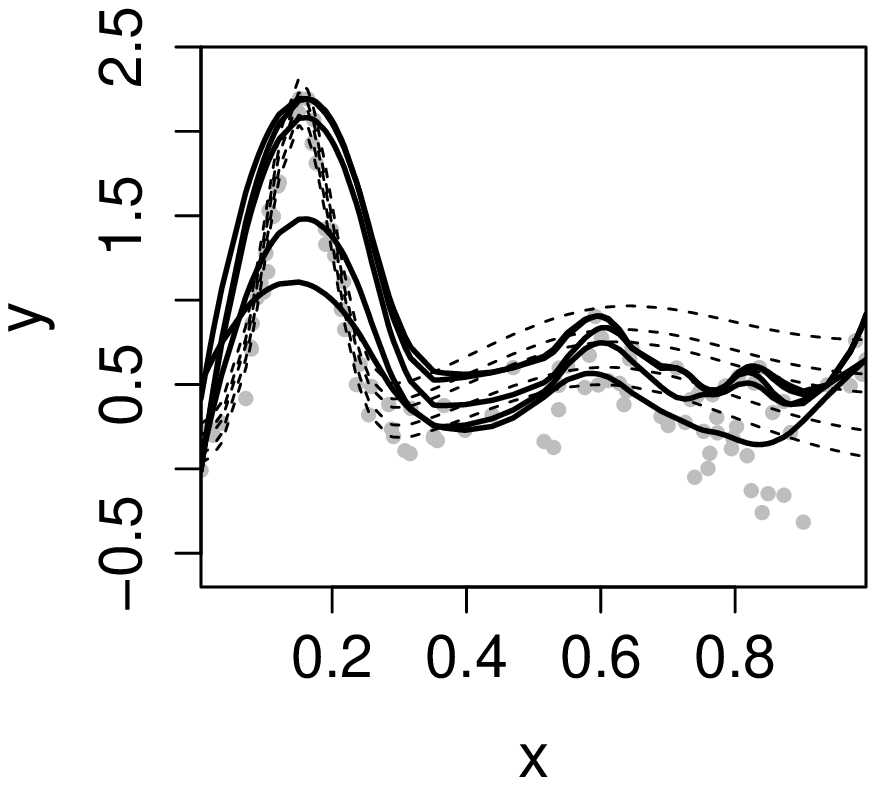}}{\hspace{0.3cm} COBs}
    \stackunder[5pt]{\includegraphics[width=4cm,height=4cm,angle=-90]{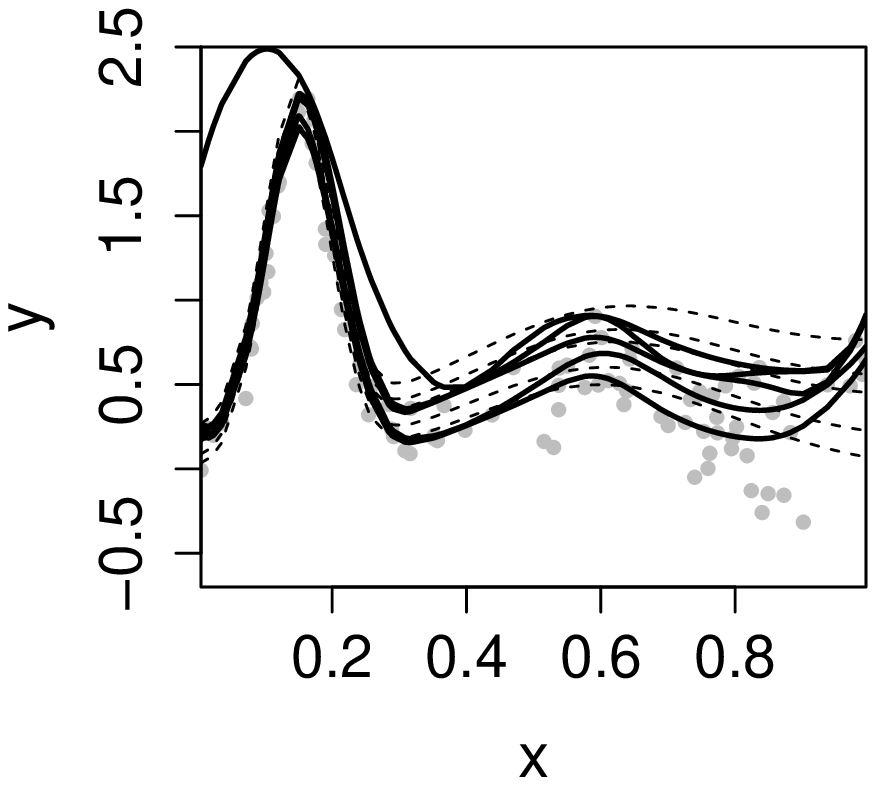}}{\hspace{0.3cm} GCQR}
    \stackunder[5pt]{\includegraphics[width=4cm,height=4cm,angle=-90]{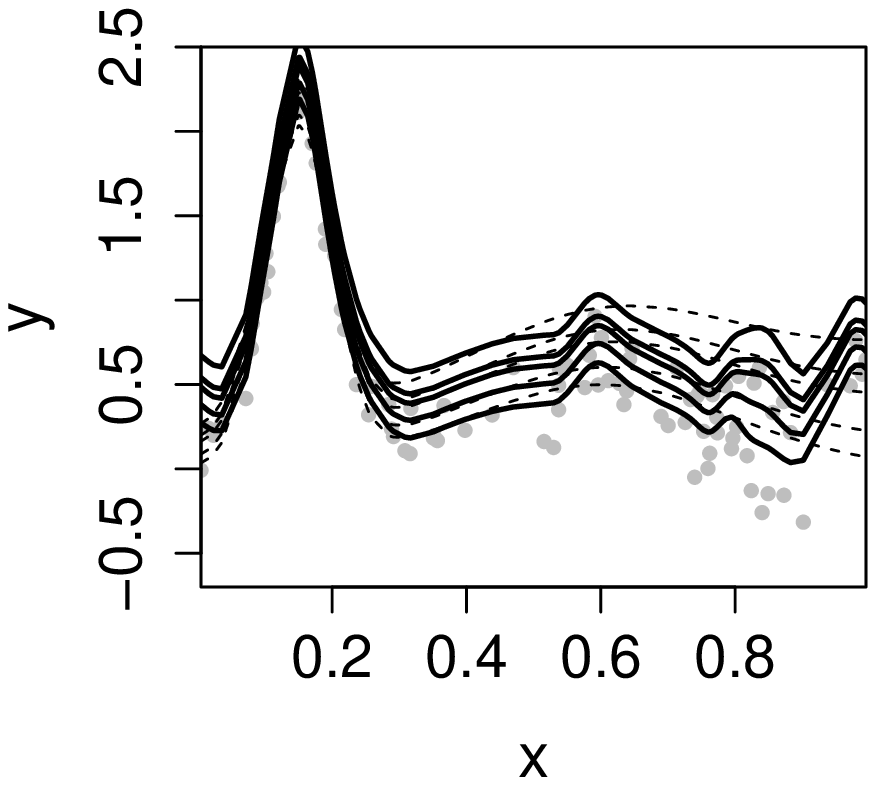}}{\hspace{0.3cm} QRJ}
    
    \caption{Estimated quantile curves at $\tau=0.5,0.7,0.9,0.95,0.99$ for one sample from Design 3 (first row) and Design 4 (second row). Dashed lines are true quantile functions, and solid lines are quantile estimates from pyramid quantile penalised spline (PQPS), constrained B-splines (COBs), growth chart regression quantiles (GCRQ) and joint estimation of linear quantile planes (QRJ).}
    %and RQ (bottom row, Koenker (2005)) at $\tau=0.25,0.5,0.75$. Right column zoom in the initial interval (2 to 12ms).
    \label{fig:des34}
\end{figure}

Simulation Designs 3 and 4 are adapted from \citeN{Smith1996}. Here we consider a sample of $n=100$ observations coming from 
\begin{description}
 \item [Design 3.] $y=\phi(x,0.15,0.1^2)/4 + \phi(x, 0.6,0.2^2)/4 + \epsilon$  ;
 \item [Design 4.] $y=\phi(x,0.15,0.05^2)/4 + \phi(x, 0.6,0.2^2)/4 + \epsilon$  ;
\end{description}
where $\phi(x,\mu, \sigma^2)$ denotes the value at $x$ of the Normal density ${\cal N}(\mu,\sigma^2)$, $x$ is an observation of $X \sim U(0,1)$ and errors are heteroscedastic $\epsilon \sim N(0, (0.1 + x/10 + x^2/10)^2)$. While Design 3 showcases a slow varying quantile function, Design 4 presents quantile curves with degree of smoothness changing abruptly with $x$. This feature can easily be incorporated into our modelling framework by specifying a fat tailed distribution for the random effects parameters $\mathbf{u}$ (Equation \ref{eq:mucenterb}), which are responsible for controlling the smoothness of the fit at different $x$. Thus, for Design 4, we assume $\mathbf{u} \sim \text{Cauchy}(\mathbf{0}, \sigma^2_u \mathbf{I})$. One sample from each design is illustrated in Figure \ref{fig:des34}, as well as the quantile fittings from PQPS, COBs, GCQR and QRJ, all using B-splines with $20$ internal knots.

From Design 4 in Figure \ref{fig:des34}, we observe that COBs quantile estimates are oversmoothed for $x<0.4$ and undersmoothed for $x>0.4$, as curve smoothness is assumed fixed, and an overall average smoothness is naturally not suitable here. In addition, quantile curves are dissociated from each other for COBs and GCQR, which clearly jeopardises the estimation of extreme quantiles. Notably, PQPS again exhibits the best model fit among the different methods. RMISE and $95\%$ coverage probabilities, based on $200$ simulated datasets, are presented in Tables \ref{tab:des3} and \ref{tab:des4}.

From Tables \ref{tab:des3} and \ref{tab:des4}, we conclude that PQPS has significantly lower RMISE for both designs and all quantile levels. COBs errors are very high for Design 4, when compared to the other methods, due to the rigid smoothness assumption. In addition, for both simulation designs, COBs coverages probabilities are well below nominal level for all quantiles. A drastic decrease is also observed for GCQR coverage probabilities at the tails. PQPS coverage is slightly under $95\%$ for Design 4, but results are still better than the other approaches.

\begin{table}[!ht] \centering 
  \caption{RMISE ($\times 100$) and $95$\% coverage probabilities for Design 3} 
  \label{tab:des3} 
\begin{tabular}{@{\extracolsep{5pt}} D{.}{.}{-2} D{.}{.}{-2} D{.}{.}{-2} D{.}{.}{-2} D{.}{.}{-2} D{.}{.}{-2} } 
\\[-1.8ex]\hline 
\hline \\[-1.8ex] 
\multicolumn{1}{c}{} & \multicolumn{1}{c}{$0.5$} & \multicolumn{1}{c}{$0.7$} & \multicolumn{1}{c}{$0.9$} & \multicolumn{1}{c}{$0.95$} & \multicolumn{1}{c}{$0.99$} \\ 
\hline \\[-1.4ex]
\multicolumn{1}{l}{\textit{RMISE}} &  &  &  &  &    \\[0.2ex]
\multicolumn{1}{l}{\hspace{0.3cm} PQPS} & 5.2 & 5.6 & 7.3 & 8.4 & 10.4  \\ 
\multicolumn{1}{l}{\hspace{0.3cm} COBs} & 7.2 & 7.8 & 9.6 & 11.4 & 18.6  \\ 
\multicolumn{1}{l}{\hspace{0.3cm} GCQR} & 6.6 & 7.0 & 9.0 & 10.9 & 18.0 \\ 
\multicolumn{1}{l}{\hspace{0.3cm} QRJ}  & 8.7 & 9.1 & 10.8 & 12.0 & 16.3  \\ [1ex]
\multicolumn{1}{l}{\textit{Coverage}} &  &  &  &  &     \\[0.2ex]
\multicolumn{1}{l}{\hspace{0.3cm} PQPS} & 0.97 & 0.96 & 0.96 & 0.96 & 0.96  \\ 
\multicolumn{1}{l}{\hspace{0.3cm} COBs} & 0.55 & 0.53 & 0.46 & 0.24 & 0.00 \\ 
\multicolumn{1}{l}{\hspace{0.3cm} GCQR} & 0.95 & 0.93 & 0.89 & 0.79 & 0.60  \\ 
\multicolumn{1}{l}{\hspace{0.3cm} QRJ}  & 0.94 & 0.93 & 0.91 & 0.90 & 0.88  \\
\hline \\[-1.8ex] 
\end{tabular} 
\end{table}

\begin{table}[!ht] \centering 
  \caption{RMISE ($\times 100$) and $95$\% coverage probabilities for Design 4} 
  \label{tab:des4} 
\begin{tabular}{@{\extracolsep{5pt}} D{.}{.}{-2} D{.}{.}{-2} D{.}{.}{-2} D{.}{.}{-2} D{.}{.}{-2} D{.}{.}{-2} } 
\\[-1.8ex]\hline 
\hline \\[-1.8ex] 
\multicolumn{1}{c}{} & \multicolumn{1}{c}{$0.5$} & \multicolumn{1}{c}{$0.7$} & \multicolumn{1}{c}{$0.9$} & \multicolumn{1}{c}{$0.95$} & \multicolumn{1}{c}{$0.99$} \\ 
\hline \\[-1.4ex]
\multicolumn{1}{l}{\textit{RMISE}} &  &  &  &  &    \\[0.2ex]
\multicolumn{1}{l}{\hspace{0.3cm} PQPS} & 8.1 & 8.4 & 9.9 & 10.9 & 12.9  \\ 
\multicolumn{1}{l}{\hspace{0.3cm} COBs} & 28.2 & 28.0 & 37.0 & 40.3 & 43.4  \\ 
\multicolumn{1}{l}{\hspace{0.3cm} GCQR} & 8.1 & 8.4 & 10.8 & 14.8 & 52.0 \\ 
\multicolumn{1}{l}{\hspace{0.3cm} QRJ} & 8.9 & 9.4 & 11.3 & 12.6 & 16.7  \\ [1ex]
\multicolumn{1}{l}{\textit{Coverage}} &  &  &  &  &    \\[0.2ex] 
\multicolumn{1}{l}{\hspace{0.3cm} PQPS} & 0.88 & 0.88 & 0.89 & 0.90 & 0.91  \\ 
\multicolumn{1}{l}{\hspace{0.3cm} COBs} & 0.48 & 0.46 & 0.32 & 0.24 & 0.00 \\ 
\multicolumn{1}{l}{\hspace{0.3cm} GCQR} & 0.97 & 0.96 & 0.88 & 0.81 & 0.55  \\ 
\multicolumn{1}{l}{\hspace{0.3cm} QRJ}  & 0.93 & 0.92 & 0.90 & 0.89 & 0.88  \\
\hline \\[-1.8ex] 
\end{tabular} 
\end{table} 

In conclusion, pyramid quantile penalised spline provides a robust quantile estimate, with significantly smaller errors than all investigated methods, and also better coverages and model fit. {In particular, the method does not drastically deteriorate in the tails. 
This appears to be a characteristic of the proposed procedure which adopts simultaneous fit, not only considering the noncrossing constraint.}

\section{Real examples} \label{sec:realex}

\subsection{Motorcycle data set}\label{sec:motorcycle}
Here we analyse the prominent motorcycle dataset (\citeNP{Silverman85}). The dataset is obtained from an experiment on the efficacy of crash helmets, and contains 133 observations of head acceleration (in g) as a function of time since impact (in milliseconds). As interest lies in describing the acceleration curve, and particularly the behaviour at the tails of the distribution, quantile regression modelling is an appealing technique which has been repeatedly considered here (see \shortciteNP{Koenker2005}, \shortciteNP{ChenYu2009}, \shortciteNP{Pratesi09} and \shortciteNP{Yanan2012}). However, all aforementioned works estimate each quantile level individually, and although quantile ordering could be imposed in a post-processing step to correct the crossing, the separate fits loose borrowing information among the quantile levels and the curves are overall dissociated from one another, as discussed on the previous section.

\begin{figure}[!ht]
    \centering
    \subfloat[PQPS \label{figa}]{\includegraphics[width=7cm,height=7cm,angle=-90]{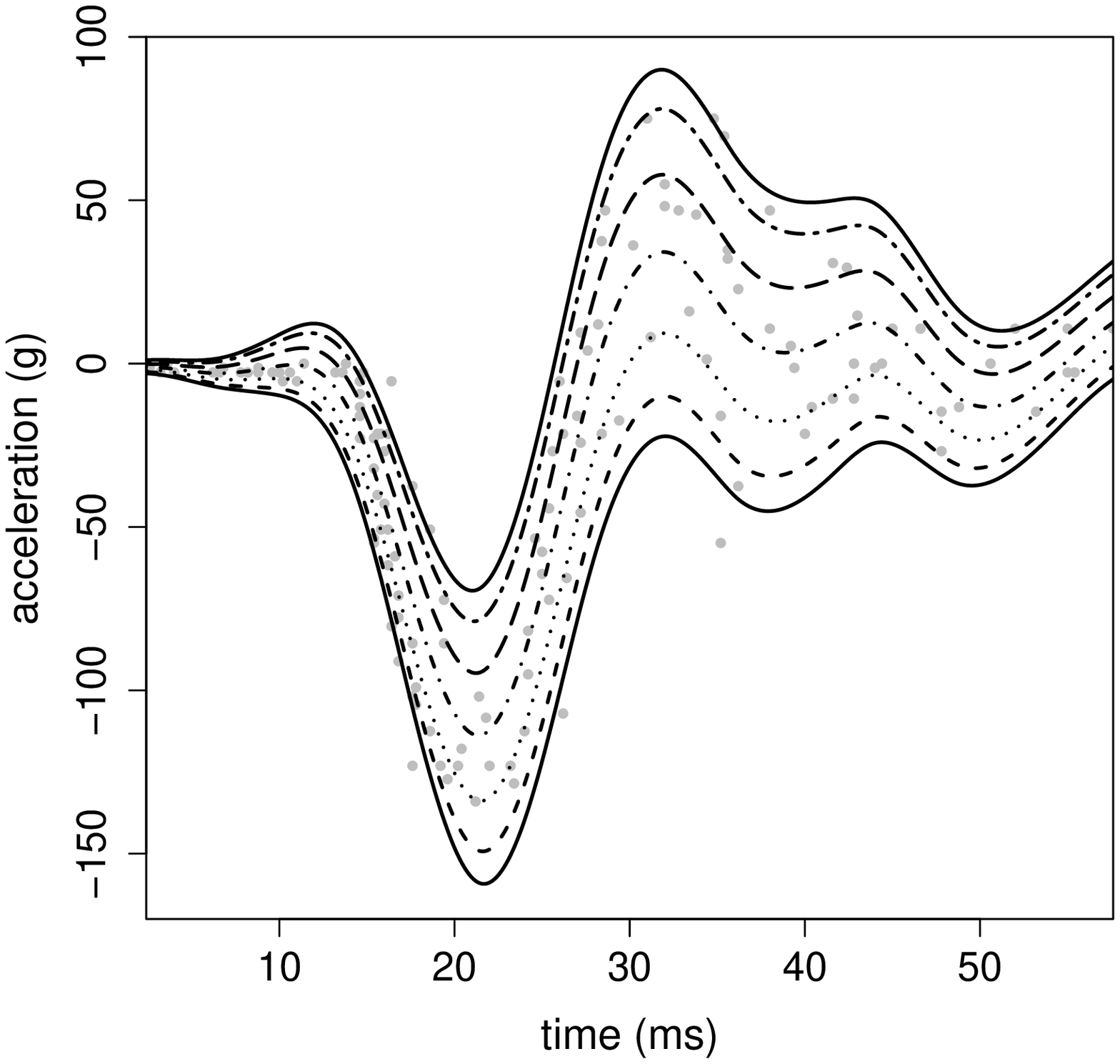}}
    \subfloat[COBS \label{figb}]{\includegraphics[width=7cm,height=7cm,angle=-90]{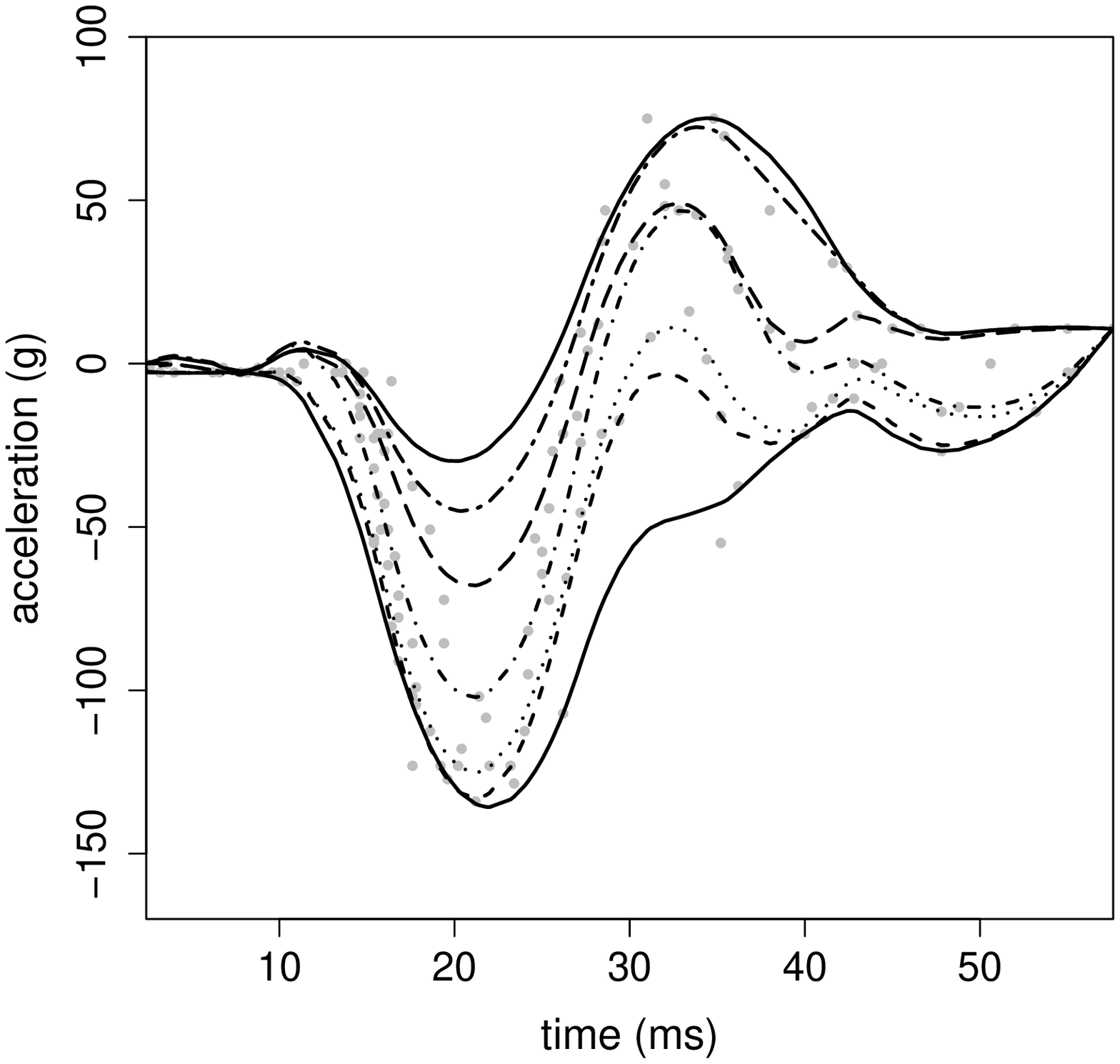}} \\
    \subfloat[GCRQ \label{figc}]{\includegraphics[width=7cm,height=7cm,angle=-90]{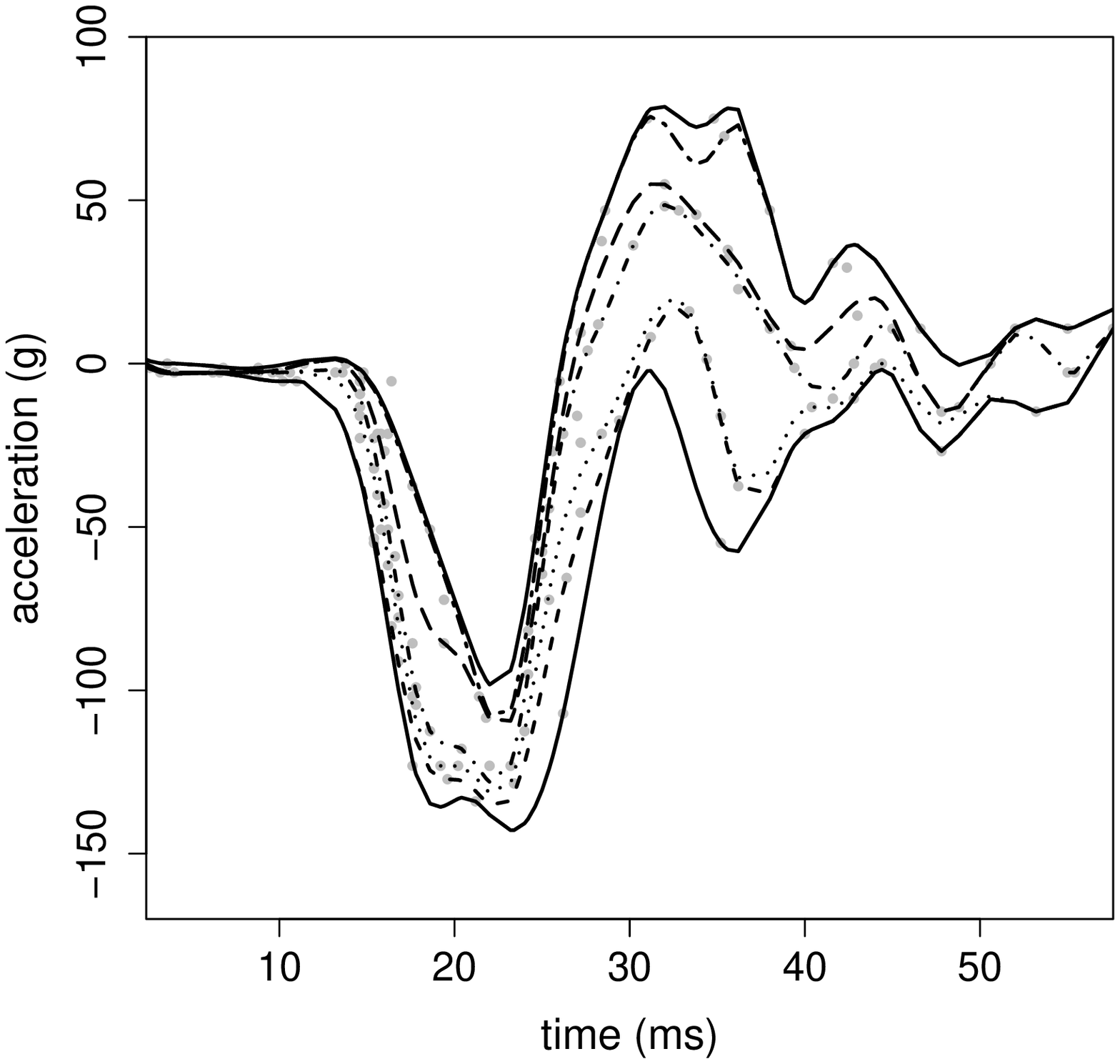}}
    \subfloat[QRJ \label{figd}]{\includegraphics[width=7cm,height=7cm,angle=-90]{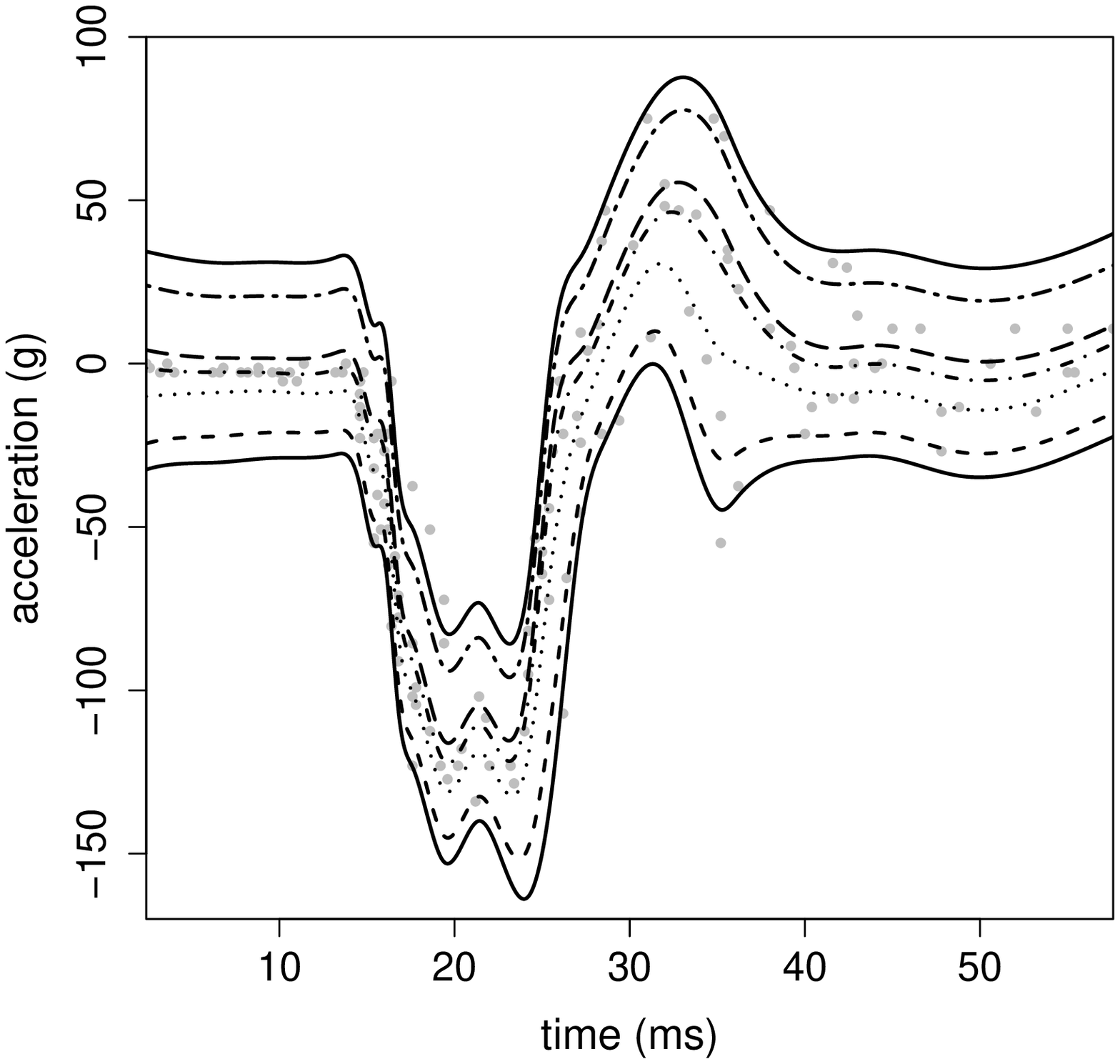}}
    \caption{Estimated quantile curves for the motorcycle dataset at $\tau=0.05,0.10,0.25,$ $0.5,0.75,0.90,0.95$ from (a) PQPS: Pyramid quantile penalised spline  (b) COBS: Constrained B-splines (c) GCRQ: Growth chart regression quantiles (d) QRJ: Joint estimation of linear quantile planes.}
    %and RQ (bottom row, Koenker (2005)) at $\tau=0.25,0.5,0.75$. Right column zoom in the initial interval (2 to 12ms).
    \label{fig:motor}
\end{figure}

We consider here the simultaneous estimation of quantile curves using smoothing B-splines with 20 internal knots. Similarly, we use an adaptive MCMC, where the covariance structure among the parameters is learnt in a first stage using $60.000$ MCMC draws and burn-in of $10.000$. 
Model parameters are estimated afterwards based on $200.000$ MCMC draws, thinning every $20$ samples, and $10.000$ burn-in. Considering that the acceleration data presents remarkably distinct patterns of variation throughout time, with a sudden change between 10ms and 30ms lying in between very smooth trajectories (see Figure \ref{fig:motor}), we again assume a Cauchy distribution for the random effects parameters, $\mathbf{u} \sim \text{Cauchy}(\mathbf{0}, \sigma^2_u \mathbf{I})$ (Equation \ref{eq:mucenterb}), in order to allow this broader range of smoothness.

Figure \ref{fig:motor}(a) shows the estimated quantile curves for PQPS {at quantile levels $\tau=0.05,$ $0.10,0.25,0.5,0.75,0.90,0.95$}. Estimated fits from alternative methods are also presented. Constrained B-splines (COBs) estimates quadratic splines (with $L_\infty$ penalty) individually for each quantile level, whereas growth chart regression quantiles (GCQR) and joint quantile regression planes (QRJ) fits jointly cubic splines, with an $L_2$ penalty and shrinkage priors, respectively.

{The proposed model for quantile regression provides a good fit to the acceleration quantile curves from the motorcycle dataset (Figure \ref{fig:motor}). PQPS quantile curves estimates are noncrossing, coherent to each other (due to the borrowing information granted by the simultaneous fit) and they nicely adapt for changes in the curve degree of smoothness. On the contrary, COBs estimates are crossing and unable to capture this varying smoothness due to the use of a global smoothness parameter, as depicted in simulation Design 4. GCRQ overall provides a reasonable fit, however there is little borrowing information among the quantile levels, so the fitted curves are dissociated from one another. As shown in the simulation studies, this is an issue for tail quantile estimation, which presents lower accuracy and coverage. The simultaneous fit provided by QRJ is noncrossing and congruent among the quantile levels, however the fit is clearly poor.}
%The simultaneous fit provided by QRJ is again an interesting candidate, offering noncrossing and congruent quantile fits, {demonstrating some of the advantages offered by Bayesian simultaneous fitting methods.}

\subsection{Immunoglobulin-G data set}
We consider the well known dataset for analysing immunodeficiency in infants. In the search for reference ranges to help diagnose infant immunodeficiency, \shortciteN{Isaac1983} measured the serum concentration of immunoglobulin-G (IgG) in 298 preschool children. As interest lies in estimating many quantiles of the response distribution, the estimates often cross. A quadratic model in age has been used previously to fit the data due to the expected smooth change of IgG with age. See   \shortciteN{rodrigues2015}, \shortciteN{Isaac1983}, \shortciteN{Yu2001} and \shortciteN{Kottas2009}. For more flexibility, we consider fitting the nonparametric spline model to this dataset.

\begin{figure}[!ht]
    \centering
    \subfloat[PQPS \label{figa1}]{\includegraphics[width=6cm,height=8cm,angle=-90]{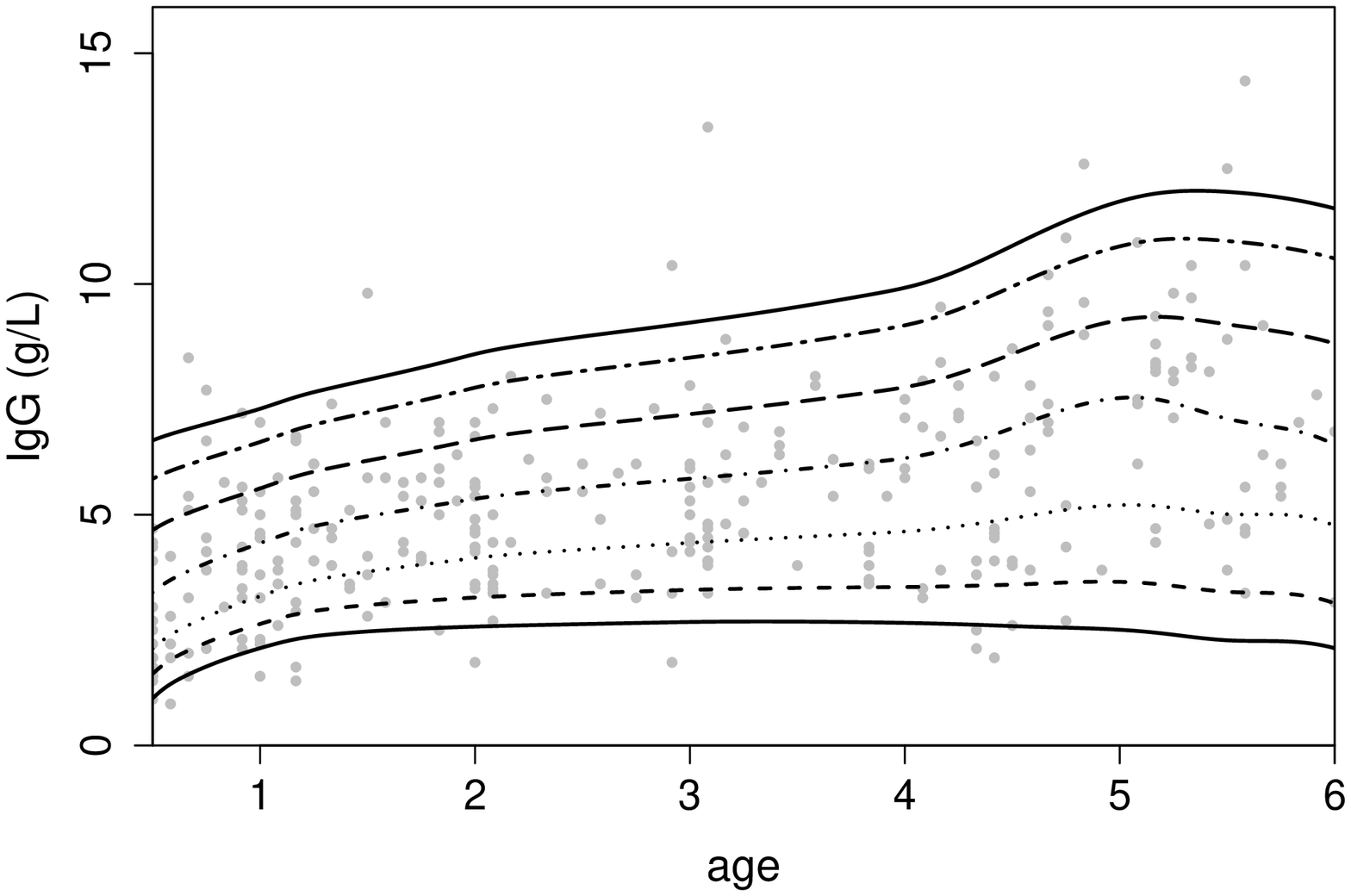}}
    \subfloat[COBS \label{figb1}]{\includegraphics[width=6cm,height=8cm,angle=-90]{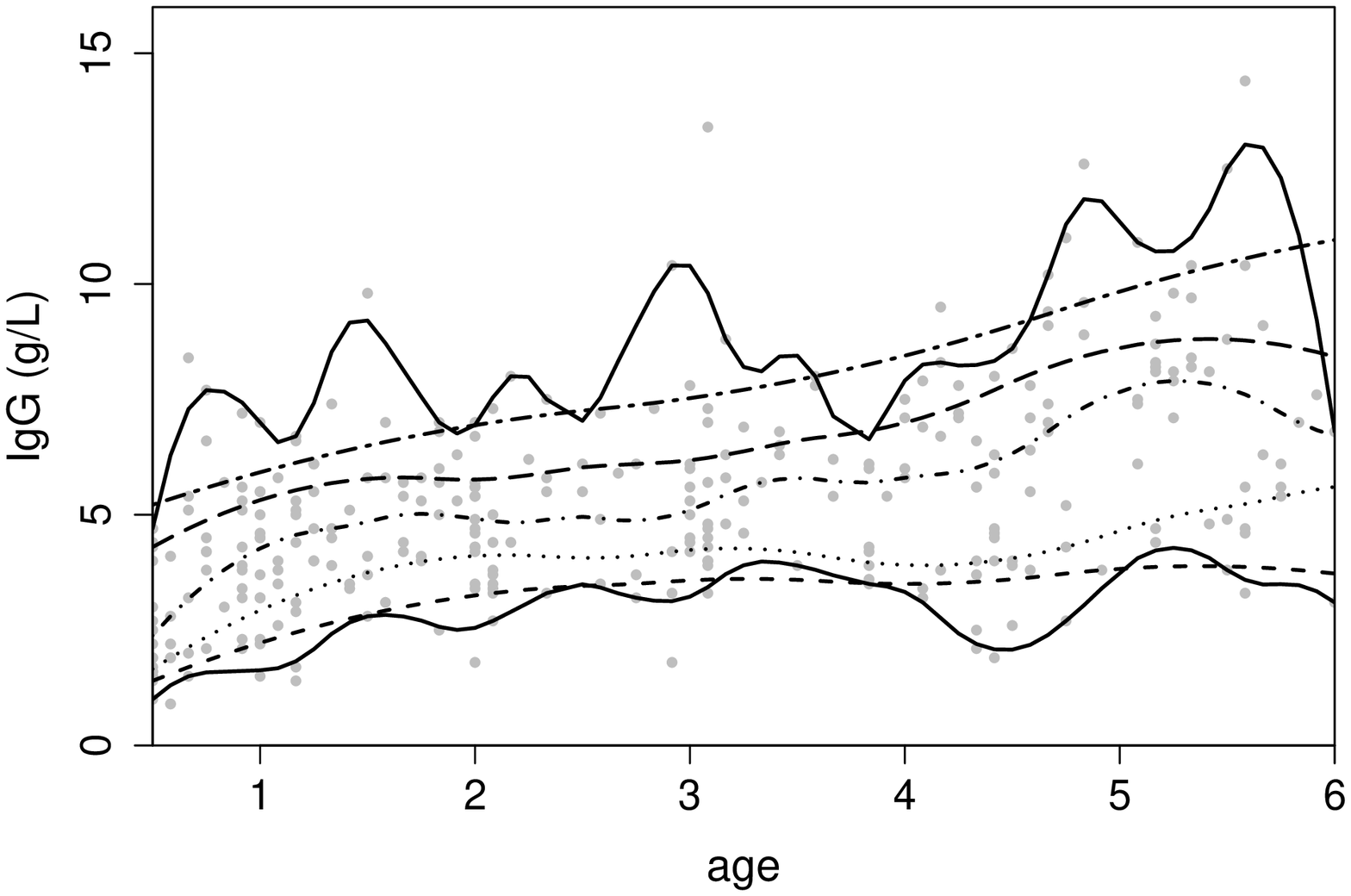}} \\
    \subfloat[GCRQ \label{figc1}]{\includegraphics[width=6cm,height=8cm,angle=-90]{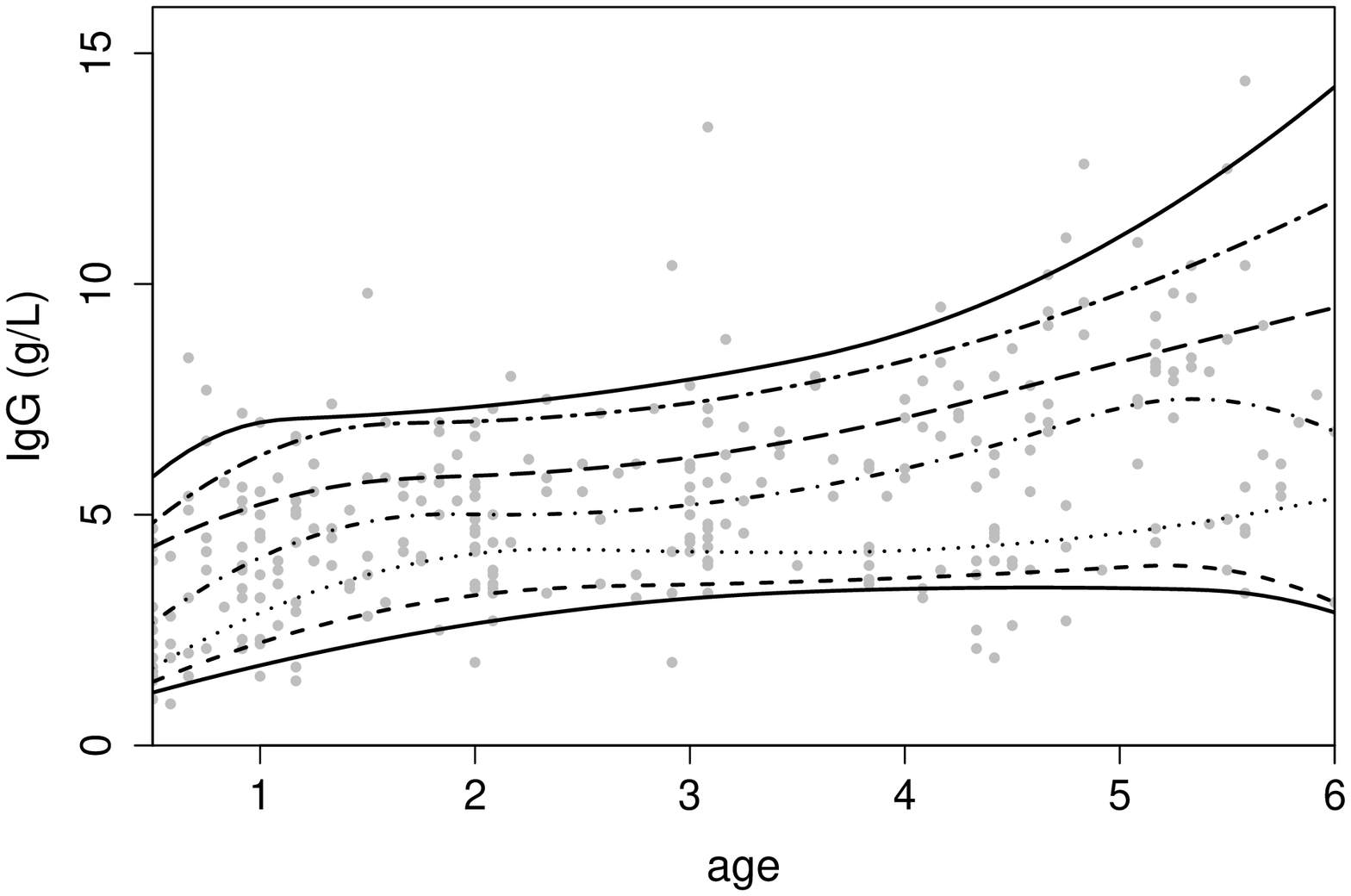}}
    \subfloat[QRJ \label{figd1}]{\includegraphics[width=6cm,height=8cm,angle=-90]{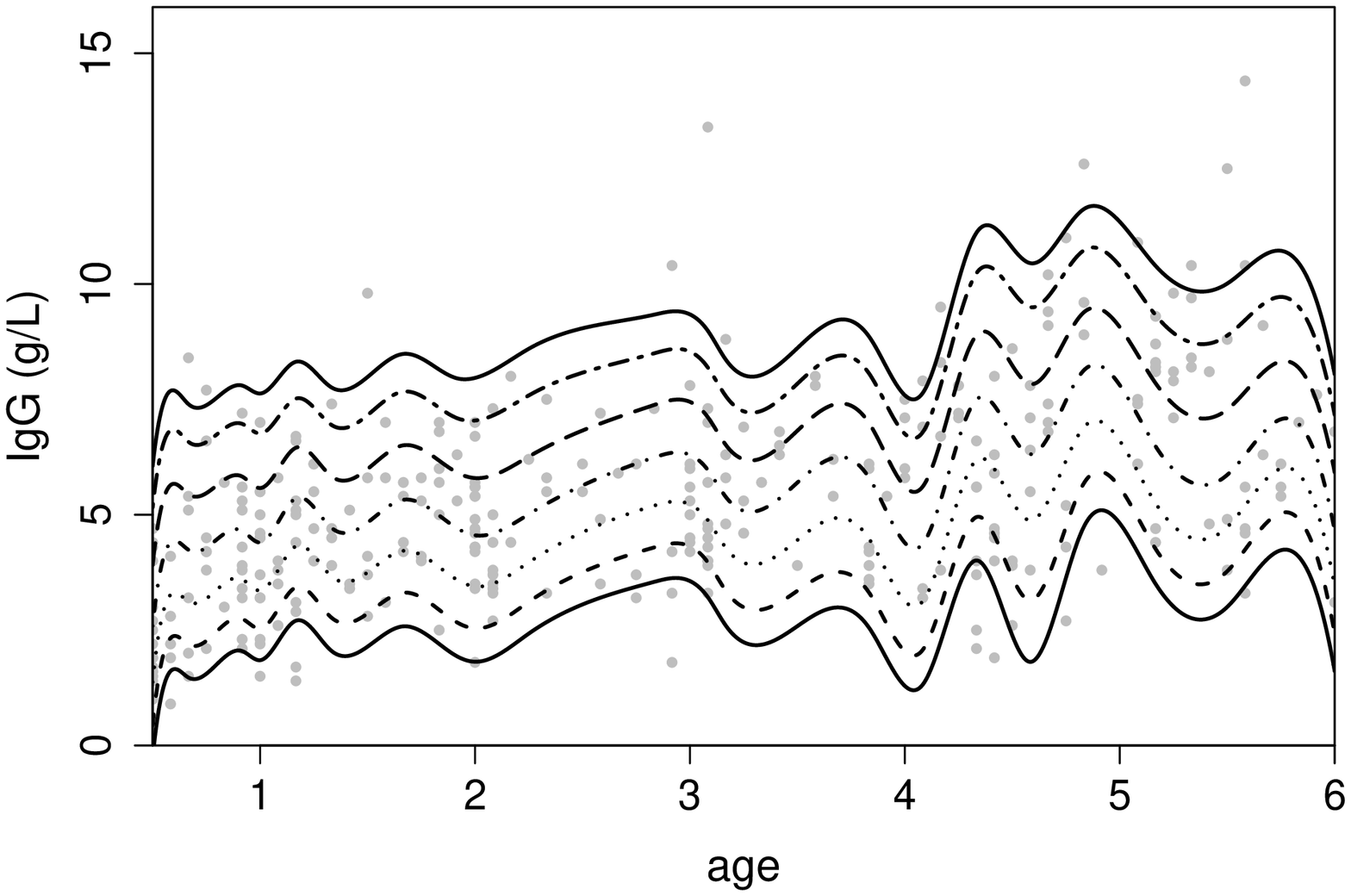}}
    \caption{{Estimated quantile curves for the IgG dataset at $\tau=0.05,0.10,0.25,$ $0.5,0.75,0.90,0.95$ from (a) PQPS: Pyramid quantile penalised spline  (b) COBS: Constrained B-splines (c) GCRQ: Growth chart regression quantiles (d) QRJ: Joint estimation of linear quantile planes.}}
    %and RQ (bottom row, Koenker (2005)) at $\tau=0.25,0.5,0.75$. Right column zoom in the initial interval (2 to 12ms).
    \label{fig:igg}
\end{figure}

We run our proposed pyramid quantile penalised splines under similar conditions as in Section \ref{sec:motorcycle}, using adaptive MCMC in a two stage procedure. The only difference 
is that here we used the Normal distribution for the random effects as in Equation (\ref{eq:mucenterb}), since the smoothness here does not vary dramatically across the covariate. Figure \ref{fig:igg} shows the fitted quantile curves at $\tau=0.05,$ $0.10,0.25,0.5,0.75,0.90,0.95$ from PQPS, COBs, GCRQ and QRJ. It is evident here that QRJ does not produce enough smoothness and provides nearly parallel curves.  Both COBs and GCRQ produced similar curves to PQPS  for all but the more extreme quantiles. However COBs shows wild behaviour 
in the two most extreme quantiles, producing crossing curves. This behaviour is also observed in earlier simulations studies for COBs. The main difference between the estimates given by GCRQ appears to be at the rightmost covariate range. Here, particularly at the higher quantile levels, GCRQ estimates suggests an increase in the IgG measurements as age increases, while both PQPS and QRJ suggest a flattening off or a small decrease. In previous simulation studies we have seen that QCRQ can perform poorly in the tails, due to the scarcity of data and not enough sharing of information across the quantiles. In this example it has lead to a rather different estimate in the higher ranges of the quantile curves.

\subsection{Lidar data set}
In this final example, we consider the famous heteroscedastic dataset of light detection and ranging (lidar) described in \shortciteN{holst96}. The dataset contains measurements of the concentration of mercury in the atmosphere, with $N=221$ observations. As concentration depends on altitude we use the distance range, {\it i.e.} the distance travelled before a laser light is reflected back to its source, as covariate. The dependent variable logratio (the logarithm of the ratio of received light from two laser sources, see \shortciteN{ruppert97} for more details) reflects the concentration of mercury, as they are an exact function of each other. This dataset has been frequently used to demonstrate smoothing for the mean regression curve.
%Since the concentration is inversely proportional to the first derivative of the dependent variable logratio (the logarithm of the ratio of received light from two laser sources, see \shortciteN{ruppert97} for more details). 

\begin{figure}[!ht]
    \centering
    \subfloat[PQPS \label{figa2}]{\includegraphics[width=6cm,height=8cm,angle=-90]{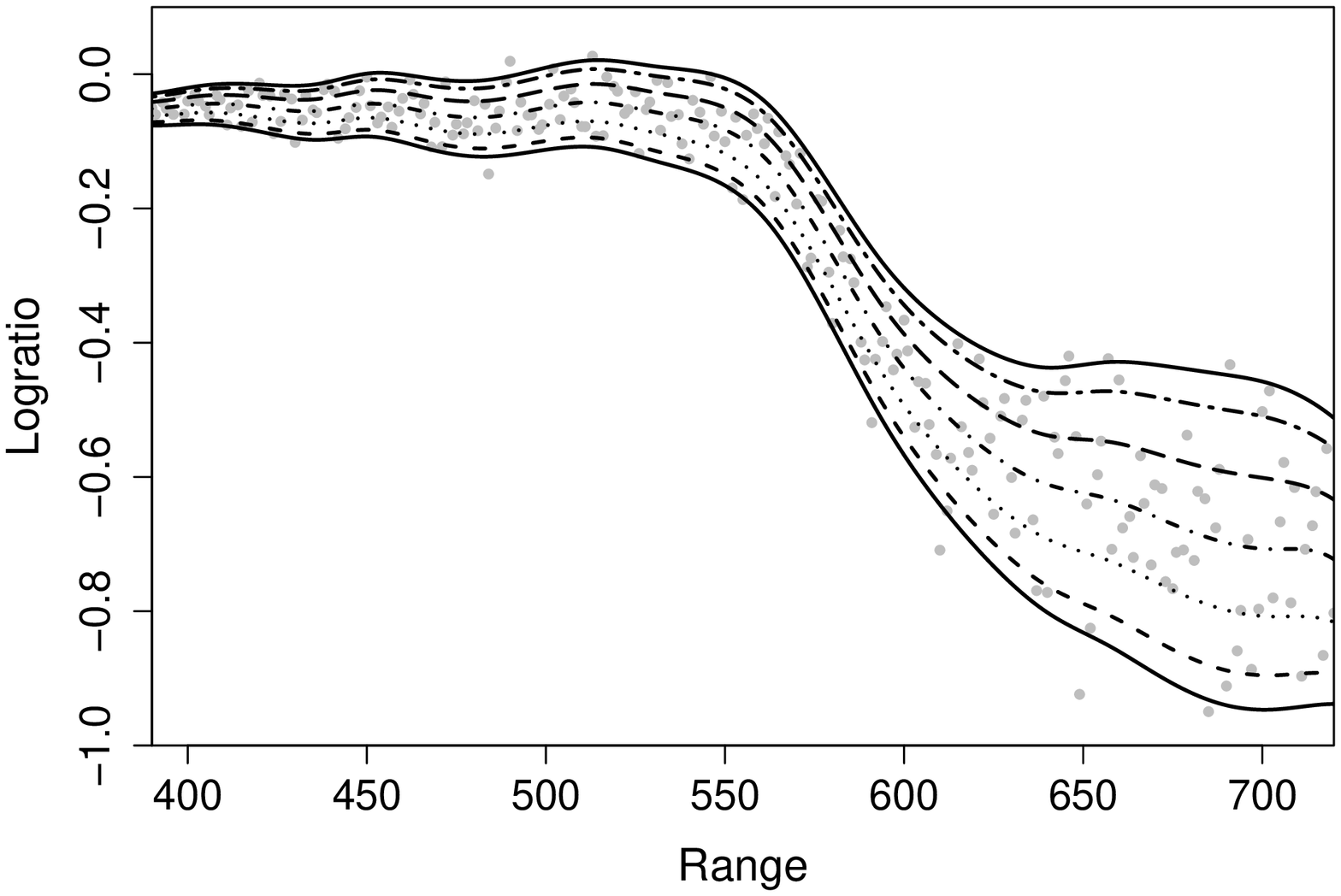}}
    \subfloat[COBS \label{figb2}]{\includegraphics[width=6cm,height=8cm,angle=-90]{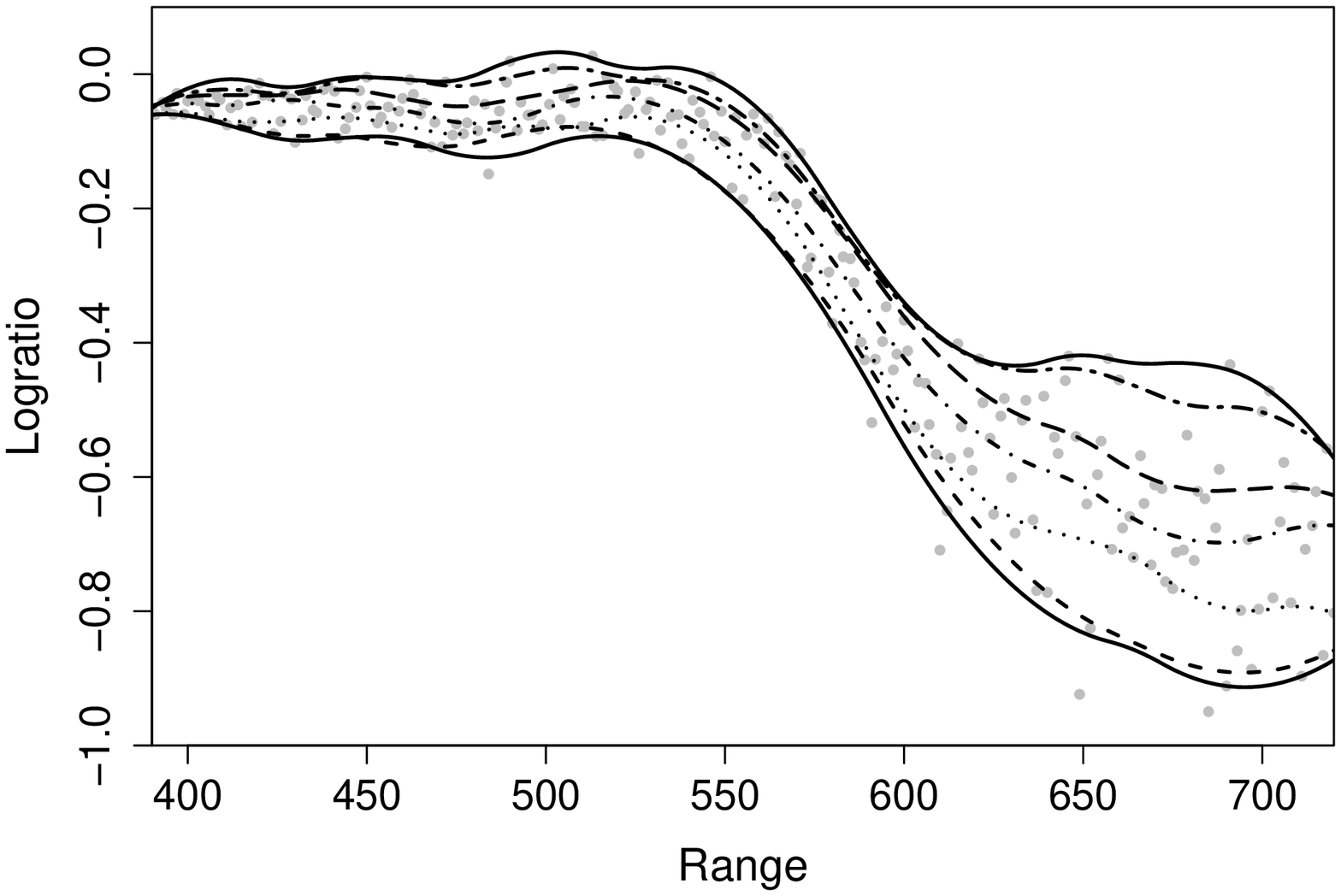}} \\
    \subfloat[GCRQ \label{figc2}]{\includegraphics[width=6cm,height=8cm,angle=-90]{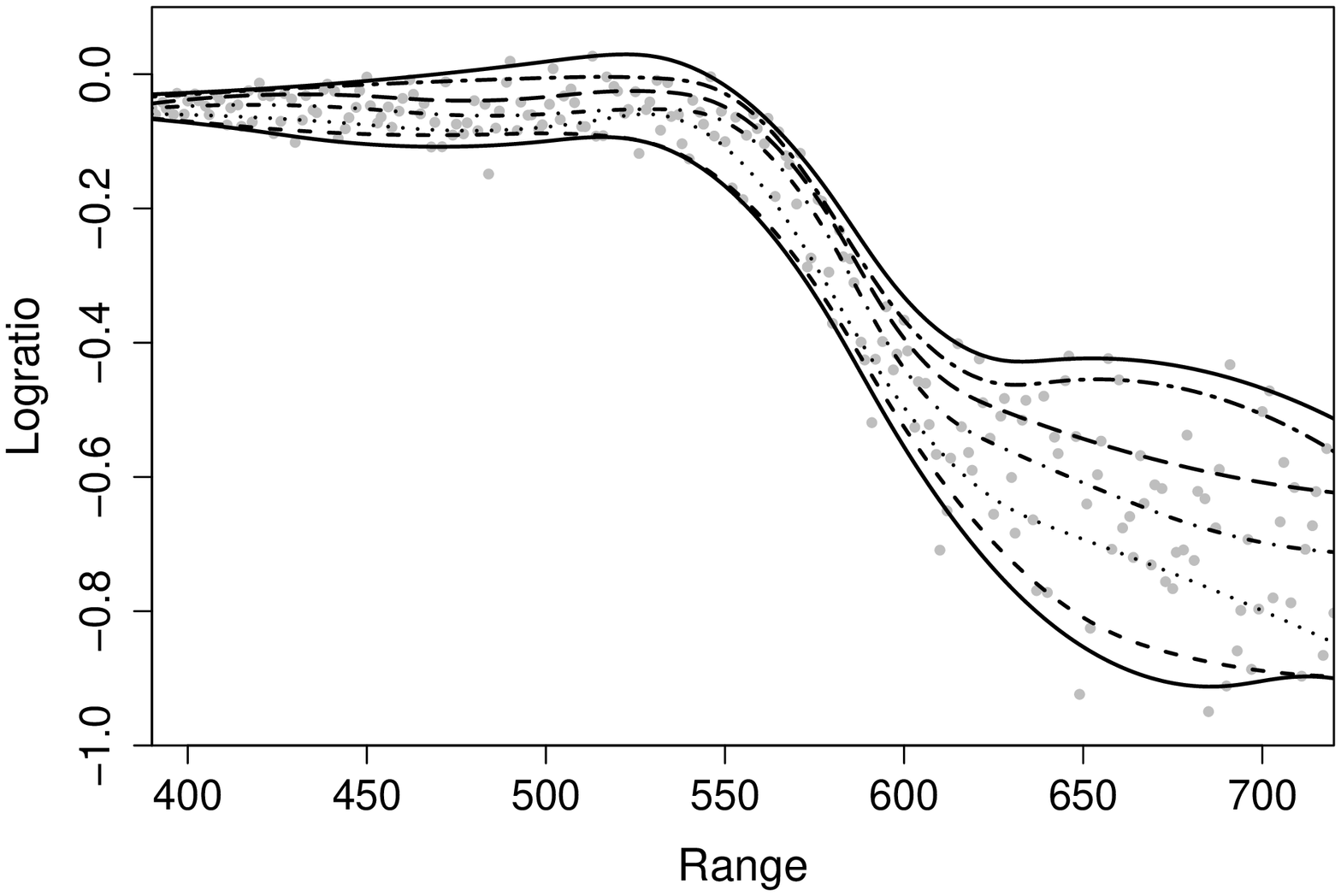}}
    \subfloat[QRJ \label{figd2}]{\includegraphics[width=6cm,height=8cm,angle=-90]{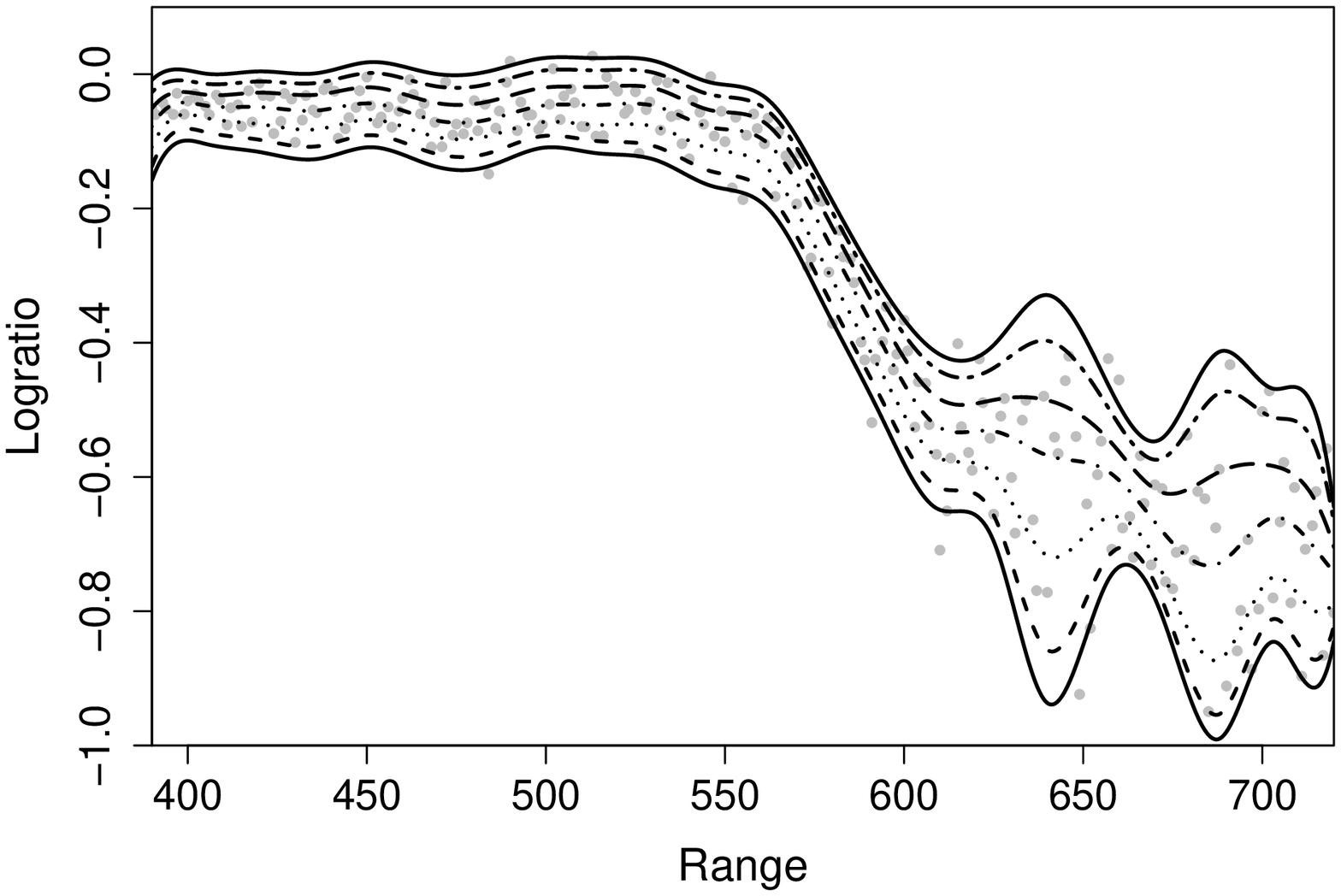}}
    \caption{{Estimated quantile curves for the LIDAR dataset at $\tau=0.05,0.10,0.25,$ $0.5,0.75,0.90,0.95$ from (a) PQPS: Pyramid quantile penalised spline  (b) COBS: Constrained B-splines (c) GCRQ: Growth chart regression quantiles (d) QRJ: Joint estimation of linear quantile planes.}}
    %and RQ (bottom row, Koenker (2005)) at $\tau=0.25,0.5,0.75$. Right column zoom in the initial interval (2 to 12ms).
    \label{fig:lidar}
\end{figure}

We again run MCMC under the same condition as in Section \ref{sec:motorcycle}, using the Cauchy distribution for the random effects parameters. Figure \ref{fig:lidar} shows the fitted
quantile curves for $\tau=0.05,0.10,0.25,$ $0.5,0.75,0.90,0.95$ using PQPS, COBS, GCRQ and QRJ. All the methods here produced visually similar results, with QRJ again not penalising enough for smoothness of the curve. While crossing does occur for this example, it is not very severe, leading to similar estimates between COBS and GCRQ. Most of the differences between the four methods of estimation appear at the leftmost of the covariate range, the COBs estimate here almost collapse to a single point, while GCRQ avoids this
problem by enforcing the non-crossing constraint. Both PQPS and GCRQ perform adequately in this example, with GCRQ producing slightly more smooth estimate than PQPS.

\section{Discussion}

In this article, we introduced a fully nonparametric quantile regression model using the quantile pyramids. Our method uses $K+4$ quantile pyramids to model
a cubic regression spline with $K$ knots, and places these pyramids on the vertices of {an optimal convex set} enclosing the predictor cloud to ensure that the fitted curves will not cross.
% proposed by \shortciteN{hjortw09}, by 
 %The pyramids are nonparametric priors on the space of quantile functions.
 %and its use in linear quantile regression was recently considered by \shortciteN{rodrigues2016}. 
 The main features of this prior are flexibility and interpretability, as it avoids strong parametric assumptions about the data, and yet provides a straightforward construction for incorporating prior information. For instance, here we considered centring the prior on the Normal distribution, nevertheless one can easily centre it in any distribution according to prior knowledge available, or incorporate different hyperparameter modelling strategies.

The nonparametric curves were modelled using cubic B-splines with a large number of knots. Smoothing was then achieved by fitting the centring mean with O'Sullivan penalised splines. More flexibility in penalisation across quantile levels can be obtained via penalties on the scaling parameter of the centring distribution, but this was found not to be necessary in our examples. We also demonstrated with simulations and real examples that local smoothness can be easily handled by assuming a heavy tailed distribution for the random effects parameters of the mixed model formulation (Section \ref{subsec:pyrmean}). 

{A feature of our modelling approach is the simultaneous estimation of quantiles, 
%The proposed modelling platform provides simultaneous quantile estimation, and as quantile levels are able to
this allows the sharing of information with each other. The advantage can be empirically observed, particularly in the tails of the distribution, where
methods which employ simultaneous fitting hold up well into the high quantiles. 
{Our procedure to find pyramid locations in high dimensions avoids the need to 
check for noncrossing constraints, allowing us to construct an effective adaptive MCMC sampling algorithm which can provide robust parameter estimates. }
% while this greatly simplifies computation, it does not greatly impact on performance.}
%\textcolor{red}{The convex hull result is important because it simplifies the procedure and allows us to build an efficient adaptive MCMC algorithm which can cope with higher dimensions. }
% more harmonious and accurate fits are delivered. In addition, by wisely selecting the pyramid locations we avoid troublesome noncrossing constraints, without imposing rigid modelling assumptions. 
Through simulation studies, and the analyse of real datasets, we showed that PQPS estimates have significantly smaller errors than the best available approaches across all quantile levels, and they also provide better coverages and model fit. 
{Extending the current framework to handle  additive modelling is trivial,  although the computational burden would be heavy}.
%Therefore, quantile pyramids for smoothing spline is a robust model for Bayesian inference on quantile regression. 
{Computational efficiency remain an issue with the pyramid quantile approach in general, recent developments in high performance software such as RStan \shortciteN{stan2017} 
and  variational Bayes approaches \shortciteN{han16}, offers some encouraging future directions
for improvements.}

\subsection*{Acknowledgements}%%
%%%%%%%%%%%%%%%%%%
%%%%%%%%%%%%%%%%%
TR is funded by CAPES Foundation via the Science Without Borders (BEX 0979/13-9). TR and YF are grateful for the support of ARC ACEMS.

\section{Appendix}

\begin{proof}[Proof of Proposition 1]
Without loss of generality, suppose that $x \in (0,1)$. Consider the cubic splines truncated power basis 
$$%\{
1, x,x^2, x^3, (x-\gamma_1)_+^3 , (x-\gamma_2)_+^3 ,\dots, (x-\gamma_K)_+^3%\}
$$ 
with $K$ knots $0<\gamma_1 < \gamma_2 < ... < \gamma_K<1$. Also, let
$${\cal X}=\{(x,x^2,x^3, (x-\gamma_1)_+^3 , \dots ,\dots, (x-\gamma_K)_+^3),x\in(0,1)\}$$ 
be the corresponding one-dimensional curve in ${\mathbb R}^{K+3}$. 
We consider finding $K+4$ locations in ${\mathbb R}^{K+3}$ whose convex hull encloses, as narrowly as possible, this curve ${\cal X}$. The following steps describe the selection of the tightest region using tangent planes.

First, we consider the first three elements of the basis, $1$, $x$ and $x^2$, with corresponding curve ${\cal X}_1=\{(x,x^2),x\in (0,1)\}$ in $\mathbb{R}^2$. 
To enclose this curve we set some pyramid locations at its two extremities, say $x_1^1=(0,0)$ and $x_1^3=(1,1)$. 
For the remaining location, we note that the smallest triangle enclosing ${\cal X}_1$ is obtained when its sides are tangent to ${\cal X}_1$ at $x_1^1$ and $x_1^3$. 
Calculating the intersection of the two tangent lines to the curve at $x_1^1$ and $x_1^3$ and we get $x_1^2=(\frac{1}{2},0)$. 
See Figure \ref{proofa}. 

\begin{figure}[!ht]
   \centering
%    \subfloat[${\cal X}_1=\{(x,x^2),x\in (0,1)\}$ \label{proofa}]{\includegraphics[width=6.5cm,height=6.5cm,angle=-90]{proof1}} \hspace{0.5cm}
%    \subfloat[${\cal X}_2=\{(x,x^2,x^3),x\in (0,1)\}$ \label{proofb}]{\includegraphics[width=6.3cm,height=6.3cm,angle=-90]{proofd}}
    \subfloat[${\cal X}_1=\{(x,x^2),x\in (0,1)\}$ \label{proofa}]{\includegraphics[width=6.5cm,angle=0]{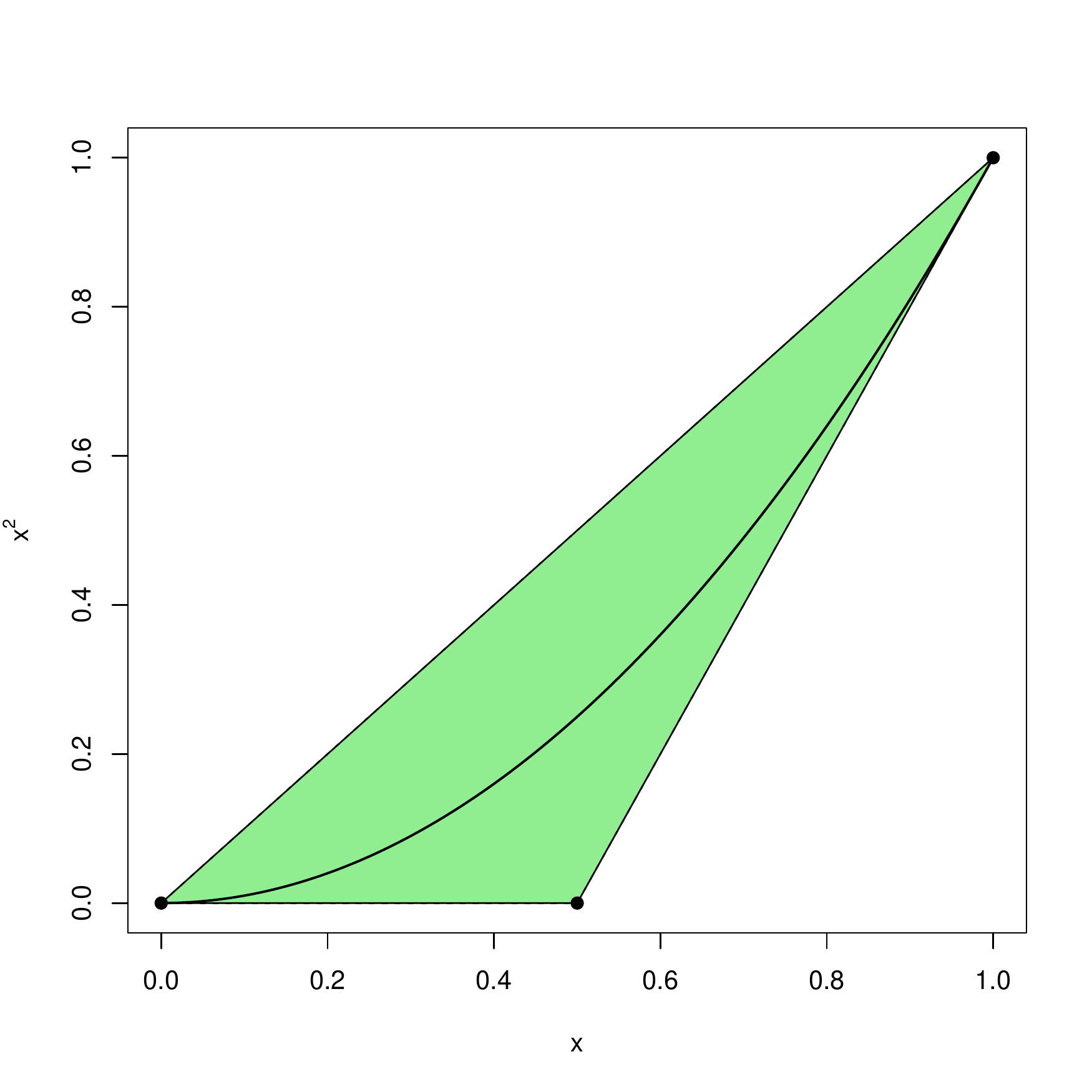}} \hspace{0.5cm}
    %\vspace{-1.5cm}
    \subfloat[${\cal X}_2=\{(x,x^2,x^3),x\in (0,1)\}$ \label{proofb}]{\includegraphics[trim=0 5cm 0 6cm, clip=true,width=7.8cm,angle=0]{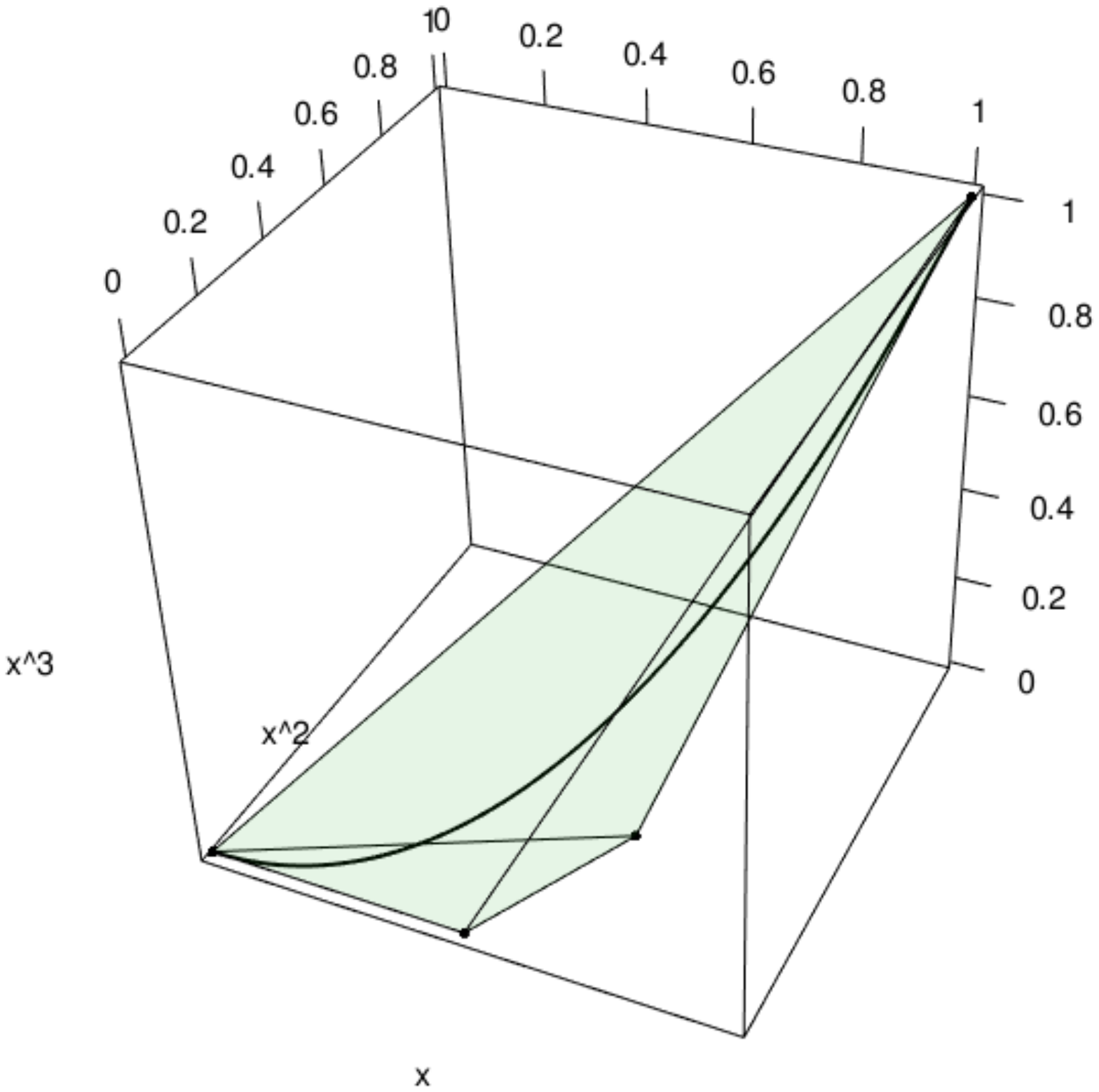}}
    %\vspace{-1.5cm}
    \caption{Pyramid locations for the basis $\{1,x,x^2\}$ (a) and $\{1,x,x^2,x^3\}$ (b). The pyramid locations are given by the dots, the corresponding convex hull are the shaded regions enclosing the curves ${\cal X}_1=\{(x,x^2),x\in (0,1)\}$ and ${\cal X}_2=\{(x,x^2,x^3),x\in (0,1)\}$.}
    \label{dessins}
\end{figure}

At the second step, we consider the first four elements of the basis and the corresponding curve in $\mathbb{R}^3$, ${\cal X}_2=\{(x,x^2,x^3),x\in (0,1)\}$.  
To enclose this curve, again, we set two pyramid locations at its extremities,  $x_2^1=(0,0,0)$ and $x_2^4=(1,1,1)$. 
Let $(x,y,z)$ be the Cartesian coordinates in  $\mathbb{R}^3$. 
Since the derivative of the curve at $x_2^1$ is $(1,0,0)'$ we know that the two remaining pyramid locations are on the plane  $\{z=0\}$. 
Since  the mapping of ${\cal X}_2$ onto this plane is ${\cal X}_1$, we re-use the pyramid location defined at first step, $x_2^2=(\frac{1}{2},0,0)$. 
For the last pyramid location, we calculate the intersection of the line given by the derivative of the curve at $x_2^4$ with the plane $\{z=0\}$. 
The line has equation $\{(x,y,z)'=(1,1,1)'+t(1,2,3)', t\in \mathbb{R}\}$, thus we get $x_2^3=(\frac{2}{3},\frac{1}{3},0)$. See Figure \ref{proofb}.

For the next step, we introduce in the basis the truncated cubic function $(x-\gamma_1)_+^3$ for a knot $0<\gamma_1<1$. To enclose the corresponding curve ${\cal X}_3=\{(x,x^2,x^3,(x-\gamma_1)_+^3),x\in (0,1)\}$, we proceed as before: two pyramid locations are set at the extremities of the curve, $x_3^1=(0,0,0,0)$ and $x_3^5=(1,1,1,(1-\gamma_1)^3)$, and we re-use the pyramid locations previously defined, now $x_3^2=(\frac{1}{2},0,0,0)$ and $x_3^3=(\frac{2}{3},\frac{1}{3},0,0)$. 
The only new pyramid location to calculate lies at the intersection of the line given by the derivative of the curve at $x_3^5$ with the hyperplane  $\{(x,y,z,0), (x,y,z)\in\mathbb{R}^3\}$ of $\mathbb{R}^4$. The line has equation $\{(x,y,z,w)'=(1,1,1,(1-\gamma_1)^3)'+t(1,2,3,3(1-\gamma_1)^2)', t\in \mathbb{R}\}$, and setting the last coordinate to $0$ we get $x_3^4=(\frac{2+\gamma_1}{3},\frac{1+2\gamma_1}{3},\gamma_1,0)$.

New truncated cubic functions can be added to the basis by induction. Suppose we want to add to the basis $1, x, x^2, (x-\gamma_1)_+^3, \dots, (x-\gamma_{K-1})_+^3$ the function $(x-\gamma_K)_+^3$, we then re-use the pyramid locations previously defined, and calculate the only new pyramid location. For that, we first determine the line given by the derivative of ${\cal X}$ at $x^{K+4}=(1,1,1,(1-\gamma_1)^3,...,(1-\gamma_K)^3)$, %line defined by 
$$
\left\{(1+t,1+2t,1+3t,(1-\gamma_1)^3+3t(1-\gamma_1)^2,\dots,(1-\gamma_K)^3+3t(1-\gamma_K)^2), t\in \mathbb{R} \right\},
$$
then setting its last coordinate to $0$, we get
\begin{equation*}
\begin{array}{ll}
x^{K+3}=&\left(\frac{2+\gamma_K}{3},\frac{1+2\gamma_K}{3},\gamma_K,(\gamma_K-\gamma_1)(1-\gamma_1)^2,(\gamma_K-\gamma_2)(1-\gamma_2)^2,\dots, \right.\\
&\left. (\gamma_K-\gamma_{K-1})(1-\gamma_{K-1})^2,0 \right).
\end{array}
\end{equation*}

\end{proof}

\bibliographystyle{chicago}
\bibliography{splines-paper}

\begin{thebibliography}{}

\bibitem[\protect\citeauthoryear{Bondell, Reich, and Wang}{Bondell
  et~al.}{2010}]{Bondell2010}
Bondell, H.~D., B.~J. Reich, and H.~Wang (2010).
\newblock Noncrossing quantile regression curve estimation.
\newblock {\em Biometrika\/}~{\em 97\/}(4), 825--838.

\bibitem[\protect\citeauthoryear{Bosch, Ye, and Woodworth}{Bosch
  et~al.}{1995}]{BOSCH95}
Bosch, R.~J., Y.~Ye, and G.~G. Woodworth (1995).
\newblock A convergent algorithm for quantile regression with smoothing
  splines.
\newblock {\em Computational Statistics \& Data Analysis\/}~{\em 19\/}(6), 613
  -- 630.

\bibitem[\protect\citeauthoryear{Cade, Terrell, and Schroeder}{Cade
  et~al.}{1999}]{ecol99}
Cade, B.~S., J.~W. Terrell, and R.~L. Schroeder (1999).
\newblock Estimating effects of limiting factors with regression quantiles.
\newblock {\em Ecology\/}~{\em 80\/}(1), 311--323.

\bibitem[\protect\citeauthoryear{Chen and Yu}{Chen and Yu}{2009}]{ChenYu2009}
Chen, C. and K.~Yu (2009).
\newblock Automatic {B}ayesian quantile regression curve fitting.
\newblock {\em Statistics and Computing\/}~{\em 19}, 271--281.

\bibitem[\protect\citeauthoryear{Chernozhukov, Fernandez-Val, and
  Galichon}{Chernozhukov et~al.}{2009}]{chernozhukovfg09}
Chernozhukov, V., I.~Fernandez-Val, and A.~Galichon (2009).
\newblock Improving point and interval estimators of monotone functions by
  rearrangement.
\newblock {\em Biometrika\/}~{\em 96}, 559--575.

\bibitem[\protect\citeauthoryear{Dette and Volgushev}{Dette and
  Volgushev}{2008}]{dettev08}
Dette, H. and S.~Volgushev (2008).
\newblock Non-crossing non-parametric estimates of quantile curves.
\newblock {\em Journal of Royal Statistical Society B\/}~{\em 70}, 609--627.

\bibitem[\protect\citeauthoryear{Dortet-Bernadet and Fan}{Dortet-Bernadet and
  Fan}{2012}]{Yanan2012}
Dortet-Bernadet, J.-L. and Y.~Fan (2012).
\newblock On {B}ayesian quantile regression curve fitting via auxiliary
  variables.
\newblock arXiv:1202.5883v1[stat.ME].

\bibitem[\protect\citeauthoryear{Fang, Chen, and He}{Fang
  et~al.}{2015}]{fengch2015}
Fang, Y., Y.~Chen, and X.~He (2015).
\newblock Bayesian quantile regression with approximate likelihood.
\newblock {\em Bernoulli\/}~{\em 21\/}(2), 832--580.

\bibitem[\protect\citeauthoryear{Fitzenberger, Koenker, and
  Machado}{Fitzenberger et~al.}{2002}]{bookQR}
Fitzenberger, B., R.~Koenker, and J.~A.~F. Machado (2002).
\newblock {\em Economic Applications of Quantile Regression}.
\newblock Physica.

\bibitem[\protect\citeauthoryear{Garthwaite, Fan, and Sisson}{Garthwaite
  et~al.}{2016}]{garthwaitefs11}
Garthwaite, P.~H., Y.~Fan, and S.~A. Sisson (2016).
\newblock Adaptive optimal scaling of metropolis-hastings algorithms using the
  robbins-monro process.
\newblock {\em Communications in Statistics - Theory and Methods\/}~{\em 45},
  5098--5111.

\bibitem[\protect\citeauthoryear{Han, Liao, Dunson, and Carin}{Han
  et~al.}{2016}]{han16}
Han, S., X.~Liao, D.~B. Dunson, and L.~Carin (2016).
\newblock Variational gaussian copula inference.
\newblock In {\em 19th International Conference on Artificial Intelligence and
  Statistics}, Volume~51, pp.\  829 -- 838.

\bibitem[\protect\citeauthoryear{Hastie and Tibshirani}{Hastie and
  Tibshirani}{1990}]{hastie+t90}
Hastie, T.~J. and R.~J. Tibshirani (1990).
\newblock {\em Generalised additive models}.
\newblock Chapman and Hall, London.

\bibitem[\protect\citeauthoryear{He}{He}{1997}]{he97}
He, X. (1997).
\newblock Quantile curves without crossing.
\newblock {\em American Statistician\/}~{\em 51}, 186--192.

\bibitem[\protect\citeauthoryear{He and Ng}{He and Ng}{1999}]{He1999}
He, X. and P.~Ng (1999).
\newblock Cobs: Qualitatively constrained smoothing via linear programming.
\newblock {\em Computational Statistics\/}~{\em 14\/}(3), 315--337.

\bibitem[\protect\citeauthoryear{Hjort and Walker}{Hjort and
  Walker}{2009}]{hjortw09}
Hjort, N.~L. and S.~G. Walker (2009).
\newblock Quantile pyramids for {B}ayesian nonparametrics.
\newblock {\em Annals of Statistics\/}~{\em 37\/}(1), 105--131.

\bibitem[\protect\citeauthoryear{Holst, Hossjer, Bjorklund, Ragnarson, and
  Edner}{Holst et~al.}{1996}]{holst96}
Holst, U., O.~Hossjer, C.~Bjorklund, P.~Ragnarson, and H.~Edner (1996).
\newblock Locally weighted least squares kernel regression and statistical
  evaluation of lidar measurements.
\newblock {\em Environmetrics\/}~{\em 7}, 401--416.

\bibitem[\protect\citeauthoryear{Isaacs, Altman, Tidmarsh, Valman, and
  Webster}{Isaacs et~al.}{1983}]{Isaac1983}
Isaacs, D., D.~G. Altman, C.~E. Tidmarsh, H.~B. Valman, and A.~D.~B. Webster
  (1983).
\newblock Serum immunoglobulin concentration in preschool children measured by
  laser nephelometry: reference ranges for {IgG}, {IgA}, {IgM}.
\newblock {\em Journal of Clinical Pathology\/}~{\em 36}, 1193--1196.

\bibitem[\protect\citeauthoryear{Koenker}{Koenker}{2005}]{Koenker2005}
Koenker, R. (2005).
\newblock {\em Quantile regression}, Volume~38 of {\em Econometric Society
  Monographs}.
\newblock Cambridge: Cambridge University Press.

\bibitem[\protect\citeauthoryear{Koenker and Bassett}{Koenker and
  Bassett}{1978}]{KoenkerBasset1978}
Koenker, R. and J.~Bassett, Gilbert (1978).
\newblock Regression quantiles.
\newblock {\em Econometrica\/}~{\em 46\/}(1), 33--50.

\bibitem[\protect\citeauthoryear{Koenker, Ng, and Portnoy}{Koenker
  et~al.}{1994}]{koenker94}
Koenker, R., P.~Ng, and S.~Portnoy (1994).
\newblock Quantile smoothing splines.
\newblock {\em Biometrika\/}~{\em 81\/}(4), 673.

\bibitem[\protect\citeauthoryear{Kottas and Krnjajic}{Kottas and
  Krnjajic}{2009}]{Kottas2009}
Kottas, A. and M.~Krnjajic (2009).
\newblock Bayesian semiparametric modelling in quantile regression.
\newblock {\em Scandinavian Journal of Statistics\/}~{\em 36}, 297--319.

\bibitem[\protect\citeauthoryear{Muggeo, Sciandra, Tomasello, and Calvo}{Muggeo
  et~al.}{2013}]{Muggeo2013}
Muggeo, V. M.~R., M.~Sciandra, A.~Tomasello, and S.~Calvo (2013).
\newblock Estimating growth charts via nonparametric quantile regression: a
  practical framework with application in ecology.
\newblock {\em Environmental and Ecological Statistics\/}~{\em 20\/}(4),
  519--531.

\bibitem[\protect\citeauthoryear{Ng and Maechler}{Ng and
  Maechler}{2007}]{Pin07}
Ng, P. and M.~Maechler (2007).
\newblock A fast and efficient implementation of qualitatively constrained
  quantile smoothing splines.
\newblock {\em Statistical Modelling\/}~{\em 7\/}(4), 315--328.

\bibitem[\protect\citeauthoryear{Ng and Maechler}{Ng and
  Maechler}{2015}]{COBSpackage}
Ng, P.~T. and M.~Maechler (2015).
\newblock {\em COBS -- Constrained B-splines (Sparse matrix based)}.
\newblock R package version 1.3-1.

\bibitem[\protect\citeauthoryear{Portnoy}{Portnoy}{2003}]{Portnoy03}
Portnoy, S. (2003).
\newblock Censored regression quantiles.
\newblock {\em Journal of the American Statistical Association\/}~{\em
  98\/}(464), 1001--1012.

\bibitem[\protect\citeauthoryear{Pratesi, Ranalli, and Salvati}{Pratesi
  et~al.}{2009}]{Pratesi09}
Pratesi, M., M.~G. Ranalli, and N.~Salvati (2009).
\newblock Nonparametric m-quantile regression using penalised splines.
\newblock {\em Journal of Nonparametric Statistics\/}~{\em 21\/}(3), 287--304.

\bibitem[\protect\citeauthoryear{Reich, Fuentes, and Dunson}{Reich
  et~al.}{2011}]{ReichFuentesDunson2011}
Reich, B.~J., M.~Fuentes, and D.~B. Dunson (2011).
\newblock Bayesian spatial quantile regression.
\newblock {\em Journal of the American Statistical Association\/}~{\em
  106\/}(493), 6--20.

\bibitem[\protect\citeauthoryear{Reich and Smith}{Reich and
  Smith}{2013}]{reichs13}
Reich, B.~J. and L.~B. Smith (2013).
\newblock Bayesian quantile regression for censored data.
\newblock {\em Biometrics\/}~{\em 69}, 651--660.

\bibitem[\protect\citeauthoryear{Roberts, Gelman, and Gilks}{Roberts
  et~al.}{1997}]{roberts1997}
Roberts, G.~O., A.~Gelman, and W.~R. Gilks (1997, 02).
\newblock Weak convergence and optimal scaling of random walk metropolis
  algorithms.
\newblock {\em Ann. Appl. Probab.\/}~{\em 7\/}(1), 110--120.

\bibitem[\protect\citeauthoryear{Roberts and Rosenthal}{Roberts and
  Rosenthal}{2001}]{roberts+r01}
Roberts, G.~O. and J.~S. Rosenthal (2001).
\newblock Optimal scaling for various {Metropolis-Hastings} algorithms.
\newblock {\em Statistical Science\/}~{\em 16}, 351--367.

\bibitem[\protect\citeauthoryear{Rodrigues, Dortet-Bernadet, and Fan}{Rodrigues
  et~al.}{2016}]{rodrigues2016}
Rodrigues, T., J.-L. Dortet-Bernadet, and Y.~Fan (2016).
\newblock Pyramid quantile regression.
\newblock arXiv:1606.05407 [stat.ME].

\bibitem[\protect\citeauthoryear{Rodrigues and Fan}{Rodrigues and
  Fan}{2017}]{rodrigues2015}
Rodrigues, T. and Y.~Fan (2017).
\newblock Regression adjustment for noncrossing {B}ayesian quantile regression.
\newblock {\em Journal of Computational and Graphical Statistics\/}~{\em 26},
  275--284.

\bibitem[\protect\citeauthoryear{RStan}{RStan}{2017}]{stan2017}
RStan (2017).
\newblock Stan development team, {RStan: the R interface to Stan. R package
  version 2.16.2}.

\bibitem[\protect\citeauthoryear{Ruppert, Wand, and Carroll}{Ruppert
  et~al.}{2003}]{book:semipar}
Ruppert, D., M.~Wand, and R.~Carroll (2003).
\newblock {\em Semiparametric Regression}.
\newblock Cambridge University Press.

\bibitem[\protect\citeauthoryear{Ruppert, Wand, Holst, and H{\"o}ssjer}{Ruppert
  et~al.}{1997}]{ruppert97}
Ruppert, D., M.~P. Wand, U.~Holst, and O.~H{\"o}ssjer (1997).
\newblock Local polynomial variance function estimation.
\newblock {\em Technometrics\/}~{\em 39\/}(3), 262--273.

\bibitem[\protect\citeauthoryear{Silverman}{Silverman}{1985}]{Silverman85}
Silverman, B.~W. (1985).
\newblock Some aspects of the spline smoothing approach to non-parametric
  regression curve fitting.
\newblock {\em Journal of the Royal Statistical Society. Series B
  (Methodological)\/}~{\em 47\/}(1), 1--52.

\bibitem[\protect\citeauthoryear{Smith and Kohn}{Smith and
  Kohn}{1996}]{Smith1996}
Smith, M. and R.~Kohn (1996).
\newblock Nonparametric regression using bayesian variable selection.
\newblock {\em Journal of Econometrics\/}~{\em 75\/}(2), 317 -- 343.

\bibitem[\protect\citeauthoryear{Spiriti, Eubank, Smith, and Young}{Spiriti
  et~al.}{2013}]{Spiriti13}
Spiriti, S., R.~Eubank, P.~W. Smith, and D.~Young (2013).
\newblock Knot selection for least-squares and penalized splines.
\newblock {\em Journal of Statistical Computation and Simulation\/}~{\em
  83\/}(6), 1020--1036.

\bibitem[\protect\citeauthoryear{{Statisticat} and {LLC.}}{{Statisticat} and
  {LLC.}}{2016}]{LaplacesDemon}
{Statisticat} and {LLC.} (2016).
\newblock {\em LaplacesDemon: Complete Environment for Bayesian Inference}.
\newblock R package version 16.0.1.

\bibitem[\protect\citeauthoryear{Thompson, Cai, Moyeed, Reeve, and
  Stander}{Thompson et~al.}{2010}]{Thompson2010}
Thompson, P., Y.~Cai, R.~Moyeed, D.~Reeve, and J.~Stander (2010).
\newblock Bayesian nonparametric quantile regression using splines.
\newblock {\em Computational Statistics and Data Analysis\/}~{\em 54\/}(4),
  1138 -- 1150.

\bibitem[\protect\citeauthoryear{Tokdar}{Tokdar}{2016}]{qrjoint}
Tokdar, S. (2016).
\newblock {\em qrjoint: Joint Estimation in Linear Quantile Regression}.
\newblock R package version 1.0-0.

\bibitem[\protect\citeauthoryear{Wand and Ormerod}{Wand and
  Ormerod}{2008}]{Wand2016}
Wand, M.~P. and J.~T. Ormerod (2008).
\newblock On semiparametric regression with o\'sullivan penalized splines.
\newblock {\em Australian \& New Zealand Journal of Statistics\/}~{\em
  50\/}(2), 179--198.

\bibitem[\protect\citeauthoryear{Yang and Tokdar}{Yang and
  Tokdar}{2017}]{yang2015}
Yang, Y. and S.~Tokdar (2017).
\newblock Joint estimation of quantile planes over arbitrary predictor spaces.
\newblock {\em Journal of the American Statistical Association\/}~{\em
  112\/}(1107-1120).

\bibitem[\protect\citeauthoryear{Yu and Moyeed}{Yu and
  Moyeed}{2001a}]{YuMoyeed2001}
Yu, K. and R.~A. Moyeed (2001a).
\newblock {B}ayesian quantile regression.
\newblock {\em Statist. Probab. Lett.\/}~{\em 54\/}(4), 437--447.

\bibitem[\protect\citeauthoryear{Yu and Moyeed}{Yu and Moyeed}{2001b}]{Yu2001}
Yu, K. and R.~A. Moyeed (2001b).
\newblock Bayesian quantiles regression.
\newblock {\em Statistics and Probability Letters\/}~{\em 54}, 437--447.

\end{thebibliography}

\end{document}